\documentclass[%
 reprint,
superscriptaddress,
nofootinbib,
 amsmath,amssymb,
 aps,
 prd,
floatfix,
]{revtex4-2}

\usepackage{graphicx}
\usepackage{dcolumn}
\usepackage{bm}
\usepackage{xcolor}
\usepackage[caption=false]{subfig}
\usepackage{hyperref}
\usepackage{ulem}
\usepackage{cleveref}
\usepackage{cprotect}
\usepackage{makecell}
\usepackage{booktabs}

\newcommand{\sbuaffil}{\affiliation{Department of Physics and Astronomy, Stony Brook University, Stony Brook NY 11794, USA}}
\newcommand{\ccaaffil}{\affiliation{Center for Computational Astrophysics, Flatiron Institute, New York NY 10010, USA}}
\newcommand{\cuaffil}{\affiliation{
 Department of Physics, Columbia University,
704 Pupin Hall, 538 West 120th Street, New York, New York 10027, USA
}}

\begin{document}

\title{The Ringdown of GW190521: Hints of Multiple Quasinormal Modes with a Precessional Interpretation}

\author{Harrison Siegel}
\email{hs3152@columbia.edu}
\cuaffil{}
\ccaaffil{}
\author{Maximiliano Isi}%
 \email{misi@flatironinstitute.org}
\ccaaffil{}
\author{Will M. Farr}%
 \email{will.farr@stonybrook.edu}
\ccaaffil{}
\sbuaffil{}

\begin{abstract}
GW190521 is a short-duration, low-frequency gravitational-wave signal in the LIGO-Virgo catalogue. The signal is consistent with the ringdown and possibly some of the inspiral-merger of an intermediate-mass binary black-hole coalescence. We find that previous models of the quasinormal mode spectrum in the ringdown of GW190521 give remnant mass and spin estimates which are not fully consistent with those of many inspiral-merger-ringdown waveforms. In our own analysis, we find that ringdown models which include both the angular ${l=2}$,~${m=1}$ and ${l=m=2}$ fundamental quasinormal modes are in full agreement with most inspiral-merger-ringdown waveforms, and in particular with the numerical relativity surrogate NRSur7dq4. We also find some support for including the ${l=3}$,~${m=2}$ fundamental quasinormal mode in our fits, building on Capano \textit{et al.}'s findings regarding a higher-frequency subdominant mode. We propose an interpretation of our GW190521 ringdown model that links precession to the excitation of ${l\neq m}$ quasinormal modes, but we do not rule out eccentricity or other interpretations.
\end{abstract}

\maketitle


\section{\label{sec:Introduction}Introduction}

In the theory of general relativity, a perturbed Kerr black-hole returns to equilibrium by radiating gravitational waves in a process known as the ringdown \cite{Kerr:Kerrmetric,Teukolsky:Kerrmetric,Newman:Kerr_Note,Teukolsky:1973ha, Press:1973zz, Teukolsky:1974yv, Andersson_BHPerturbationSurvey1999, secondorder_perturbations}. The ringdown emission begins with an initial burst that is quickly dominated by an infinite spectrum of discrete quasinormal modes (QNMs), which then themselves decay and give way to a power-law tail. The ringdown is expected to describe the gravitational wave emission of any remnant black hole produced by the merger of two black holes. Gravitational waves from these remnants can be observed by detectors like LIGO, Virgo, and KAGRA~\cite{AdvancedLIGOScientific:2014pky, VIRGO:2014yos, KAGRA:2020tym}. The QNM spectrum has been found to dominate the remnant's gravitational-wave signal at times after the peak of the signal's strain. The individual QNMs are essentially damped sinusoids; although their amplitudes and phases are related nontrivially to the initial conditions of the perturbation, the frequency and damping rate of every QNM depends solely on the mass and spin of the remnant black hole. Observation of the QNM spectrum emitted by a merger remnant allows for the inference of progenitor properties and validation of the Kerr metric \cite{testgr_ringdown_2,testgr_ringdownbayesian,testgr_overview}, and has recently been the subject of intense data analysis efforts \cite{IsiNoHair_GW150914,Isi_BHArea,TestingGR_LIGO_2ndCatalog,Isi_revisitGW150914,FinchMoore_GW150914, Gregorio_ringdown,Ma:2023vvr} following the dawn of gravitational-wave astronomy \cite{GW150914_detection,GWAstro_Review,intro_GWastro}.

GW190521\_030229, henceforth GW190521, is one of the more exceptional gravitational-wave signals observed so far \cite{GWTC-3_Paper, GW190521g_DiscoveryPaper, GW190521g_properties}. Due to the signal's unusual short-duration and low-frequency morphology, many interpretations of the source of GW190521 have been proposed \cite{Eccentric_RomeroShaw,Eccentric_Gayathri,Gamba_nonspinning,ProcaStar_GW190521,Estelles_GW190521Waveforms,Olsen_GW190521_Likelihood}. Under the default hypothesis that the source of GW190521 is a quasi-circular binary black-hole (BBH) coalescence, the remnant is an intermediate-mass black-hole and the ringdown comprises the majority of the observed signal.

Previous ringdown analyses of GW190521 were performed by the LIGO-Virgo collaboration (LVC)
\cite{GW190521g_DiscoveryPaper,GW190521g_properties,TestingGR_LIGO_2ndCatalog} and Capano \textit{et al.} \cite{CapanoGW190521g_330,Capano_GW195021Validation}. Both used different data analysis techniques and implemented ringdown models in line with the existing literature on non-precessing quasi-circular BBH coalescences. As we discuss in Sec.~\ref{sec:Previous_Data_Analysis} and App.~\ref{sec:Appendix_PreviousAnalysesCont}, we find that both analyses produce remnant mass and spin posteriors which are not fully consistent with the posteriors of many inspiral-merger-ringdown (IMR) waveforms including the numerical relativity (NR) surrogate NRSur7dq4 \cite{NRSur_Paper}, the preferred waveform in the GW190521 detection and properties papers \cite{GW190521g_DiscoveryPaper, GW190521g_properties}.

Most of the current IMR waveforms are only valid for quasi-circular binaries, and different waveforms parameterize the merger-ringdown signal differently~\cite{NRSur_Paper, imrphenomxphm, IMRPhenomTPHM_Paper, Ghosh:2021mrv, Brito:2018rfr}: some rely on QNM fitting formulas and inspiral-attachment conditions and assume rigid built-in relationships between the individual QNMs, while others entirely do away with an explicit damped sinusoid parameterization. By comparison, our ringdown analysis is designed to be agnostic about the amplitude and phase relationships of the QNMs, and is thus a less-constrained implementation of perturbation theory; see~\cite{AnalyzingBHRingdowns} for more details. Our fits serve as an independent check of the IMR waveforms and may be able to model physics that the IMR waveforms do not account for. The disagreement we find between posteriors of the IMR waveforms and the previous GW190521 ringdown studies raises the possibility that either some or all of these analyses have not modeled essential physics phenomena in the signal, or alternatively that these analyses may have technical systematic biases. These issues motivate our current work.

In this paper, we analyze the ringdown of GW190521 by fitting both Kerr and non-Kerr spectra of damped sinusoids to the signal in the time domain, using the \textsc{ringdown} package \cite{AnalyzingBHRingdowns,ringdown_code}. In the Kerr case, we only fit QNMs from first-order perturbation theory. In the non-Kerr case, some QNMs are allowed to independently deviate their frequency and damping rate away from Kerr values. Our primary aim is to either find a Kerr model of the QNM spectrum that gives remnant mass and spin posteriors consistent with those of IMR waveforms, in particular NRSur7dq4, or to otherwise show that an IMR-inconsistent Kerr model is preferred. In Sec.~\ref{sec:Previous_Data_Analysis} we reproduce and comment on key results of the previous GW190521 ringdown analyses, and then in Sec.~\ref{sec:Results} we perform our own fits with different sets of QNMs.

Ultimately, we find that a model which includes the angular ${l=2}$,~${m=1}$ fundamental QNM together with the ${l=m=2}$ fundamental QNM can produce remnant mass and spin posteriors consistent with NRSur7dq4 over a range of fitting times. Moreover, we also find that this model can accommodate a third QNM which is subdominant to the other two, and that we identify as the ${l=3}$,~${m=2}$ fundamental mode. We find that statistical goodness-of-fit metrics do not definitively prefer any one ringdown model. Thus, interpretation of the ringdown of GW190521 must rely on other aspects of the QNM fits such as the stability of their inferred parameters when fit over a range of start times or their consistency with the physics assumptions of our default quasi-circular BBH hypothesis. When testing general relativity by fitting non-Kerr spectra, we find a ${\sim\pm}$20\% constraint at the 90\% credible level around zero deviation from the Kerr frequency of the ${l=3}$ QNM in our model, and a less stringent constraint on the frequency of one ${l=2}$ QNM. In Sec.~\ref{sec:Discussion} we propose that the excitation of $l\neq m$ QNMs could be related to precession; since NRSur7dq4 prefers large progenitor spins in the orbital plane for GW190521, this proposal provides a cohesive physics explanation for our QNM fits. However, we do not rule out other interpretations, like eccentricity~\cite{Eccentric_Gayathri, Eccentric_RomeroShaw}. We also address the implications of our analysis for parameter estimation of BBH coalescences. We conclude in Sec.~\ref{sec:Conclusion}. In Apps.~\ref{sec:Appendix_AnalysisTechnicalChoices}--\ref{sec:Appendix_RD_TheoreticalAnalysis_Details} we discuss technical aspects of both this analysis and the previous ringdown analyses of GW190521. The data release for this paper can be found here~\cite{DataRelease}.

\subsection{\label{sec:Conventions}{Conventions}}

We will refer to each individual QNM with a sequence of three integers corresponding to the indices $lmn$. We define a ringdown model as the set of all QNMs included in a given fit, and we will refer to each model by using a comma-separated list of each included QNM enclosed within braces, e.g. \{220,~210,~320\}. The $lm$ indices denote angular content, while the $n$ index is related to radial content. QNMs that share the same $lm$ also share similar frequencies, while QNMs with the same $n$ share similar damping rates. The positive and negative values associated with a given $|m|$ encode the two polarization degrees of freedom for a given mode~\cite{AnalyzingBHRingdowns,Isi_2022mbx}.~\footnote{In the ringdown literature, the sign of $m$ is often used to distinguish retrograde and prograde QNMs---we do not adopt that convention, and instead restrict $m$ to its native role as an angular-harmonic index. Note that prograde QNMs are defined such that ${\mathrm{sgn}(m) = \mathrm{sgn}(\Re\, \tilde{\omega})}$. See discussion in \cite{AnalyzingBHRingdowns}.}
Parity-time symmetry of the Kerr metric in general relativity implies that the complex QNM frequencies obey ${\tilde{\omega}_{lmn}=-{\tilde{\omega}_{l-mn}}^*}$. This means that the temporal evolution of a given $+m$ mode is indistinguishable from the corresponding $-m$ mode when viewed from one point in the sky, and thus we will implicitly only refer to $\lvert m \rvert$. We assume the QNM parameterization described in \cite{AnalyzingBHRingdowns}.

QNM decay rates increase with increasing $n$; ${n=0}$ modes are the longest-lived so-called ``fundamental'' modes, whereas ${n>0}$ modes are the shorter-lived ``overtones".  While QNMs can have a prograde or retrograde sense, for data-driven reasons we consider only prograde QNMs here. Since the prograde 220 mode appears to be found around 70 Hz in GW190521, the retrograde $l=2$ modes would be at such low frequencies that they would likely be out of band given the shape of the LIGO-Virgo noise power spectral density; as for the retrograde $l=3$ modes, at low SNRs they are degenerate in frequency and damping rate with the prograde $l=2$ modes.

We will frequently make use of the ringdown evolution timescale ${t_M=GM/c^3}$, which is defined in units of the final remnant mass $M$ when natural units are taken such that ${G=c=1}$. For reference, we will use the median detector-frame remnant mass ${M = 258~M_\odot}$ inferred by NRSur7dq4 to set the value of $t_M$ to be 1.27 ms, and we will interchangeably refer to this value as $t_{M_\text{NRSur}}$.

Empirically, it has been found that the peak gravitational wave strain of a BBH coalescence corresponds to the earliest time at which QNM models may accurately describe the signal \cite{Giesler_ImoprtanceOvertonesRingdown,FinchMoore_PrecessingRingdown,Xiang_Ringdown,Ma:2022wpv}. The previous GW190521 analyses and our own work all use slightly different estimates of the peak strain time. The previous estimates lie within a standard deviation of our own median estimate, with one standard deviation being ${\sim 2.5~t_M}$; see Table~\ref{tab:t0_compare}, as well as Fig.~\ref{fig:SNRs_and_LOOs}. We will refer to the time at which a ringdown fit starts by either stating the GPS time $t_0$, or quoting its relative difference with respect to the median peak time as $\Delta t_0$. In figures, we will refer to the peak times of previous analyses as $t_{LVC}$~\cite{GW190521g_properties} and $t_{Capano}$~\cite{CapanoGW190521g_330}.

\begin{table}[ht]
  \centering
  \caption{Comparison of peak strain GPS time estimates (seconds) at the LIGO Hanford (H1) detector for GW190521 analyses. We use the median of the peak time distribution of ${h^2(t) \equiv \sum_{\ell m} |h_{\ell m}(t)|^2}$, obtained with NRSur7dq4 samples.}
  \label{tab:t0_compare}
  \setlength{\tabcolsep}{8pt} 
  \begin{tabular}{p{.10\linewidth}*{2}{p{.18\linewidth}}}
    \toprule
    \multicolumn{1}{p{.10\linewidth}}{\textbf{Analysis}} & \multicolumn{1}{p{.20\linewidth}}{\textbf{\makecell[l]{H1 Peak Time \\ -1242442967.0 s}}} & \multicolumn{1}{p{.18\linewidth}}{\textbf{\makecell[l]{Sky Location \\ (ra, dec)}}} \\
    \multicolumn{1}{l}{LVC \cite{GW190521g_properties}} & \multicolumn{1}{l}{0.4306} & \multicolumn{1}{l}{(0.10, -1.14)} \\
    \multicolumn{1}{l}{Capano \textit{et al.} \cite{CapanoGW190521g_330}} & \multicolumn{1}{l}{0.4259} & \multicolumn{1}{l}{(3.50, 0.73)} \\
    \multicolumn{1}{l}{Siegel \textit{et al.}} & \multicolumn{1}{l}{0.4278 $\pm$ 0.0029} & \multicolumn{1}{l}{(5.75, -0.42)} \\
    \bottomrule
  \end{tabular}
\end{table}

\section{\label{sec:Previous_Data_Analysis}Previous Analyses}

We begin by reproducing and commenting on previous analyses of the QNM spectrum of GW190521. These previous analyses were performed by the LVC \cite{GW190521g_properties} and Capano \textit{et al.}~\cite{CapanoGW190521g_330,Capano_GW195021Validation}.

\begin{figure}
    \includegraphics[width=\columnwidth]{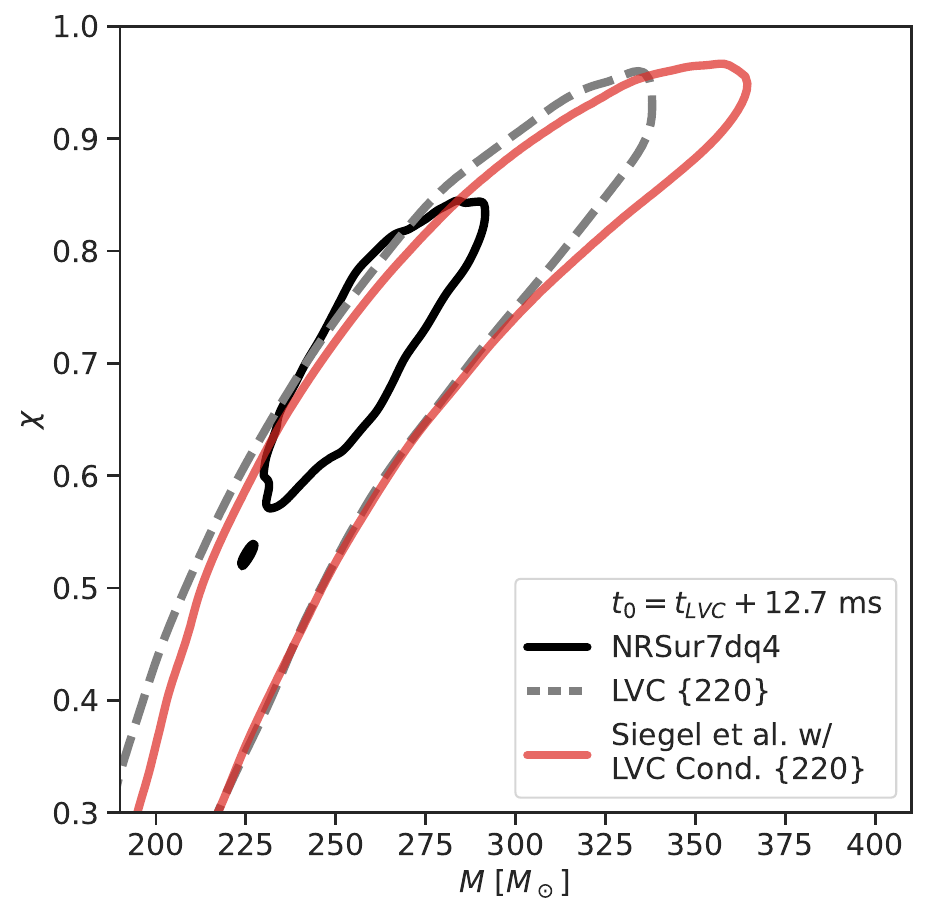}
    \caption{We attempt to reproduce the Kerr \{220\} LVC remnant mass (abscissa) and spin (ordinate) posterior from Fig.~9 of \cite{GW190521g_properties}, fitting 12.7 ms after the peak strain time reported by the LVC, and we plot 90\% credible contours. The LVC posterior is in dashed grey, and our own posterior is in red. The NRSur7dq4 fit (solid black) is from the LVC data release. Here we have implemented our own noise model but are otherwise performing similar data conditioning to that of the LVC, i.e., low- and high-pass filtering at the same frequencies, using the same duration of data segment and the same sample rate; we also use a comparable QNM amplitude prior.}
    \label{fig:Fig_LVC_10M_220_mchi}
\end{figure}

\begin{figure}
  {%
    \includegraphics[width=0.9875\columnwidth]{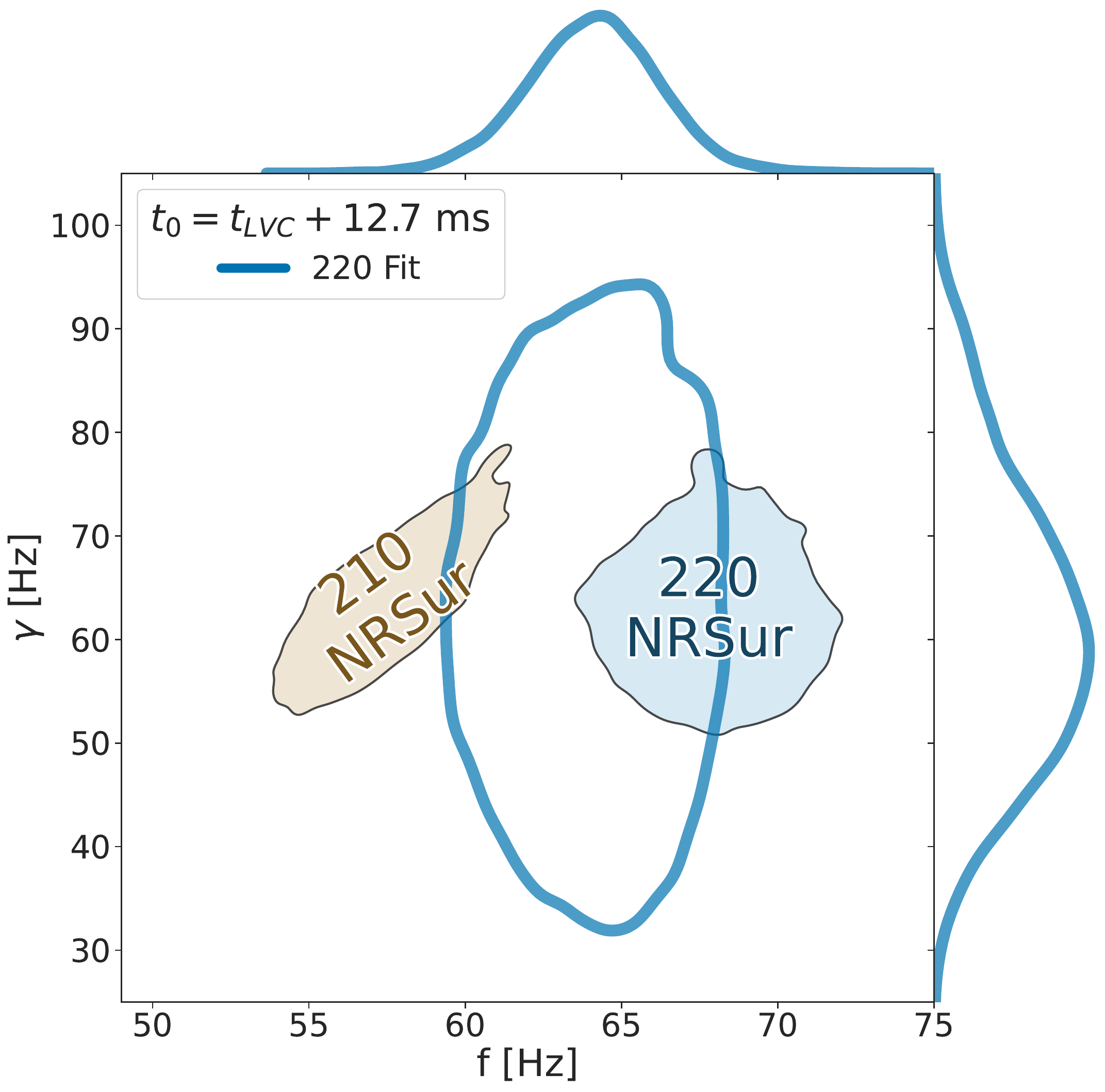}
  }\hfill
  {%
    \includegraphics[width=1.0125\columnwidth]{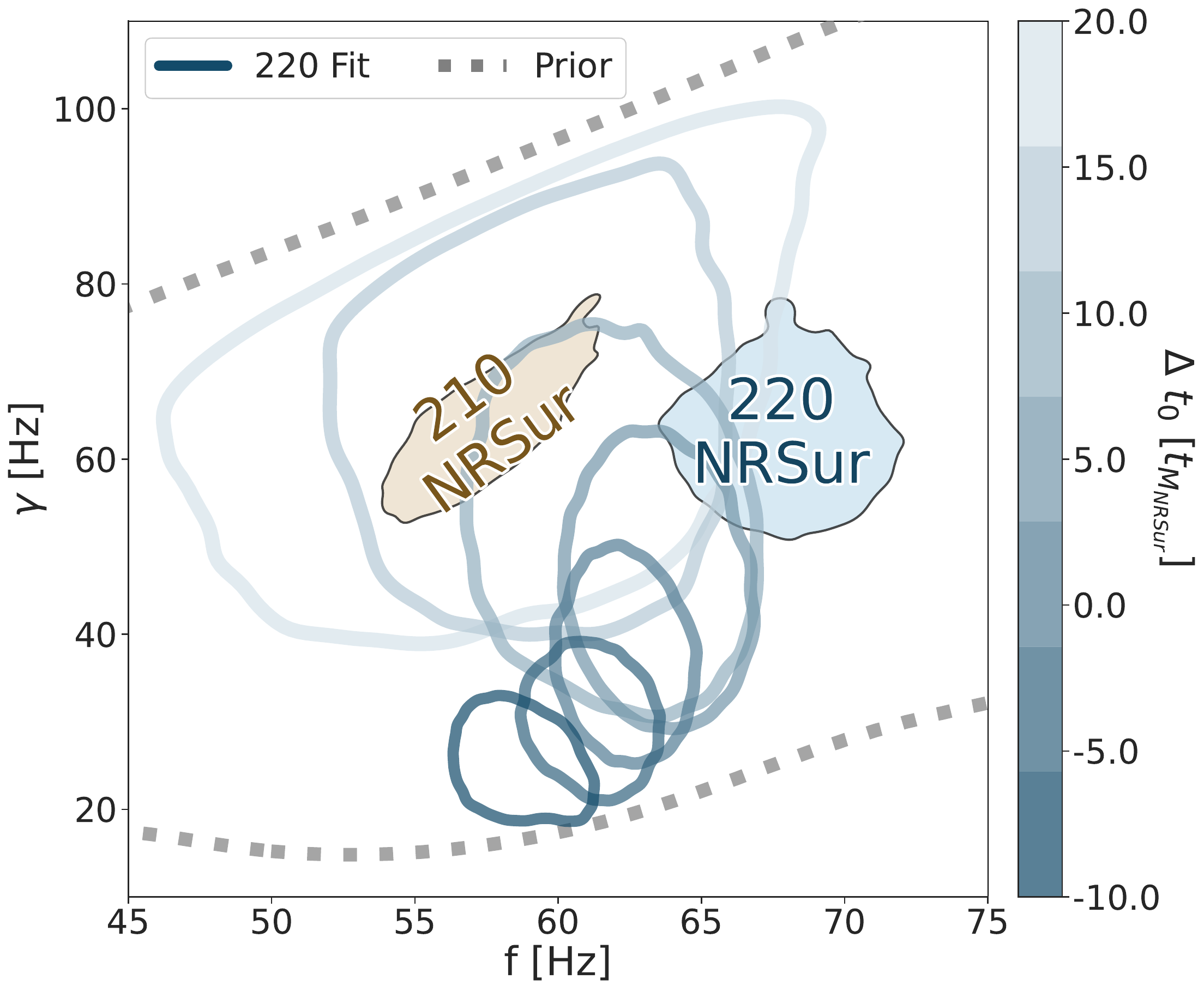}
}
    \cprotect\caption{We use remnant mass and spin samples from NRSur7dq4 along with the \textsc{qnm} package \cite{qnmpackage_Stein} to infer the frequencies and damping rates of relevant QNMs (filled contours), and we compare these to the posteriors of our QNM fits (unfilled contours). We show 90\% credible contours, with frequency and damping-rate as abscissa and ordinate respectively. In comparison with Fig.~\ref{fig:Fig_LVC_10M_220_mchi}, here it is more evident that there is tension between the \{220\} and NRSur7dq4 fits.  (Top) Our \{220\} fit at the LVC's reported time 10 $t_M$ after the peak has a frequency between the 220 and 210 frequencies of NRSur7dq4. (Bottom) We do not find a fit start time where our \{220\} fit fully agrees with the 220 of NRSur7dq4. At later times we actually find that our single-mode fit becomes consistent with the 210 of NRSur7dq4. At times past 20 $t_M$, our fit reverts to the prior. The LVC finds similar trends in their single-mode fits over time, as shown in the left plot of Fig.~9 in \cite{GW190521g_properties}, although they interpret this behavior differently.}
    \label{fig:220_fgamma}
\end{figure}

\begin{figure*}
  {\includegraphics[width=\textwidth]{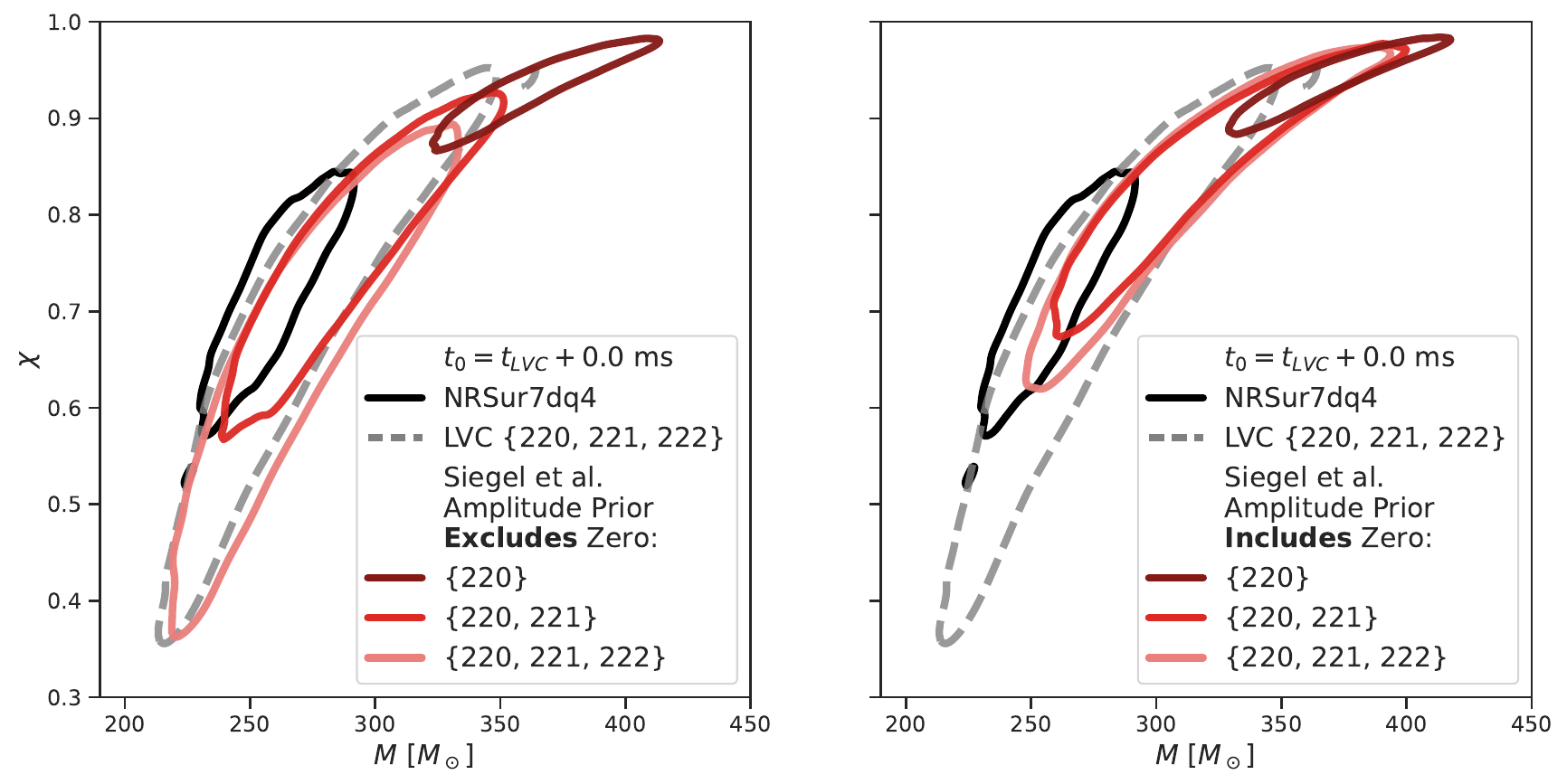}}
   \caption{We attempt to reproduce the LVC \{220,~221,~222\} remnant mass and spin posterior at their reported peak strain time. Fits with fewer overtones are shown for reference. (Left) We use a Gaussian prior on each individual QNM amplitude. This prior is similar to the triangular amplitude prior used by the LVC---crucially, neither prior has support at zero. We plot 90\% credible contours (red shades) and recover the original LVC result, up to small technical differences. (Right) We re-do the analysis, but now use a flat amplitude prior. Unlike the Gaussian prior, the flat prior supports zero amplitude. This allows for the overtone models to reduce to the \{220\} model, which increases the tension with NRSur7dq4. The nesting of these models' posteriors indicates that the overtones in these models are not strictly required for better fits to the data.}
    \label{fig:FigLVCCompare_0M}
\end{figure*}

\subsection{LVC: Analysis Overview}
The first ringdown analysis of GW190521 was performed by the LVC, using the \textsc{pyRing} code \cite{Pyring, Carullo:2019flw, IsiNoHair_GW150914, TestingGR_LIGO_2ndCatalog} to fit several distinct models consisting of damped sinusoids in the time domain. At their defined peak strain time, the LVC fit a Kerr \{220,~221,~222\} model. At ${\Delta t_0 = 12.7~\text{ms}\equiv 10~t_M}$ after the peak, the LVC fit a Kerr \{220\} model and a model with higher-order angular modes up to ${l = 4}$ and ${m= (l,~l-1)}$. The higher-order model uses amplitudes tuned as functions of the binary parameters to numerical relativity (NR) simulations of non-precessing systems \cite{LionelLondon_RD_NR_Model}. Lastly, at 6.4,~12.7,~and~19.1~ms (5,~10,~and~15~$t_M$) after the peak, the LVC fit a model consisting of a generic single damped sinusoid. In the next two sections we comment on the Kerr \{220\}, \{220,~221,~222\}, and generic damped sinusoid models; in App.~\ref{sec:Appendix_PreviousAnalysesCont} we briefly discuss the higher-order model and touch on more technical aspects of agreement between ringdown and IMR posteriors. A separate \textsc{pyRing} analysis of GW190521 with minor technical differences appears in~\cite{TestingGR_LIGO_2ndCatalog}.

\subsection{LVC: \{220\} Fit}
In Fig.~\ref{fig:Fig_LVC_10M_220_mchi} we attempt to reproduce the Kerr \{220\} fit of the LVC~\cite{GW190521g_properties}, and we recover a similar result. At first glance, the remnant mass and spin posteriors of the NRSur7dq4 and \{220\} fits seem to agree reasonably well. However, as shown in Fig.~\ref{fig:220_fgamma}, it is more evident that the \{220\} and NRSur7dq4 fits are in tension when we plot our measurements in the QNM frequency and damping rate domain. To compare our results to NRSur7dq4 in this space, we use samples of remnant mass and spin from NRSur7dq4 to estimate the frequencies and damping rates of relevant QNMs. When we fit our own \{220\} model at the time used by the LVC, the frequency of the single mode we measure lies between the 220 and 210 frequencies inferred from NRSur7dq4. Furthermore, when we perform \{220\} fits over a range of fit start times, a discrepancy is always present and at later times the single mode we fit actually becomes more consistent with the 210 of NRSur7dq4. The posteriors of our \{220\} model do not fully encompass the 220 frequency and damping rate inferred from NRSur7dq4 until ${20-30~t_M}$ after the peak, when our fits revert to the prior.

Since we use a flat remnant mass and spin prior which does not correspond to a flat QNM frequency and damping rate prior, it is worth considering the extent to which the trend over time in Fig.~\ref{fig:220_fgamma} is prior-driven. To this end, we note that the generic single damped sinusoid fits of the LVC have flat priors in frequency and damping rate, and also exhibit a similar trend (left plot of Fig.~9~in~\cite{GW190521g_properties}). One possible explanation for these trends is that the late-time GW190521 signal might be better-described by a model with more than one fundamental QNM.
 
A single-mode \{220\} fit is only expected to accurately describe the late-time ringdown signal in the special case of a non-precessing quasi-circular BBH coalescence with roughly equal-mass progenitors. For these systems, the posteriors of late-time \{220\} ringdown fits should fully overlap with the 220 mode given by IMR waveforms. This type of agreement can be seen in Fig.~5~of~\cite{PRL_TestsofGR_GW150914} for GW150914, a gravitational wave signal whose source is likely a non-precessing quasi-circular BBH. By contrast, the source of GW190521 could be precessing, and it has not been shown that the ringdowns of precessing systems are always dominated solely by the 220 at late times.

\subsection{\label{sec:222Fits}LVC: \{220,~221,~222\} Fit}

In Fig.~\ref{fig:FigLVCCompare_0M} we attempt to reproduce the LVC \{220,~221,~222\} remnant mass and spin posterior at the time of peak strain. Although we can recover a similar posterior, we find that this is contingent on the choice of QNM amplitude priors. The LVC analysis reports using a flat prior on the amplitudes of the individual left and right circular polarizations of the signal. This can be seen in Eq.~(7) of \cite{TestingGR_LIGO_2ndCatalog}, which shows the template for the QNM model of the analysis. Since this template separately adds up the amplitudes of each $\pm m$ angular index and both of these polarizations are combined in our observational models (see Sec.~\ref{sec:Conventions}), the resulting prior on the full amplitude of each damped sinusoid in the fits of the LVC is triangular instead of uniform; see Fig.~12 of \cite{Isi_2022mbx} for further details. The triangular prior prevents any mode included in the fit from having zero amplitude, forcing every mode to significantly contribute to the fit. If we instead use priors which do allow the mode amplitudes to go to zero as in the right panel of Fig.~\ref{fig:FigLVCCompare_0M}, the resulting posteriors for fits with overtones then contain the \{220\} posterior and are in greater tension with NRSur7dq4. This suggests that, while forcing overtones to contribute to fits of this signal can lead to improved qualitative IMR agreement, the overtones in these models are not required to get better fits to this data.

Defining ``good agreement'' between IMR and ringdown posteriors raises rather subtle issues. We explore these issues further in App.~\ref{sec:Appendix_PreviousAnalysesCont} and contend that even partial overlaps like the one exhibited between the LVC~\{220,~221,~222\} and NRSur7dq4 posteriors may not constitute good agreement.

\begin{figure}
    \includegraphics[width=\columnwidth]{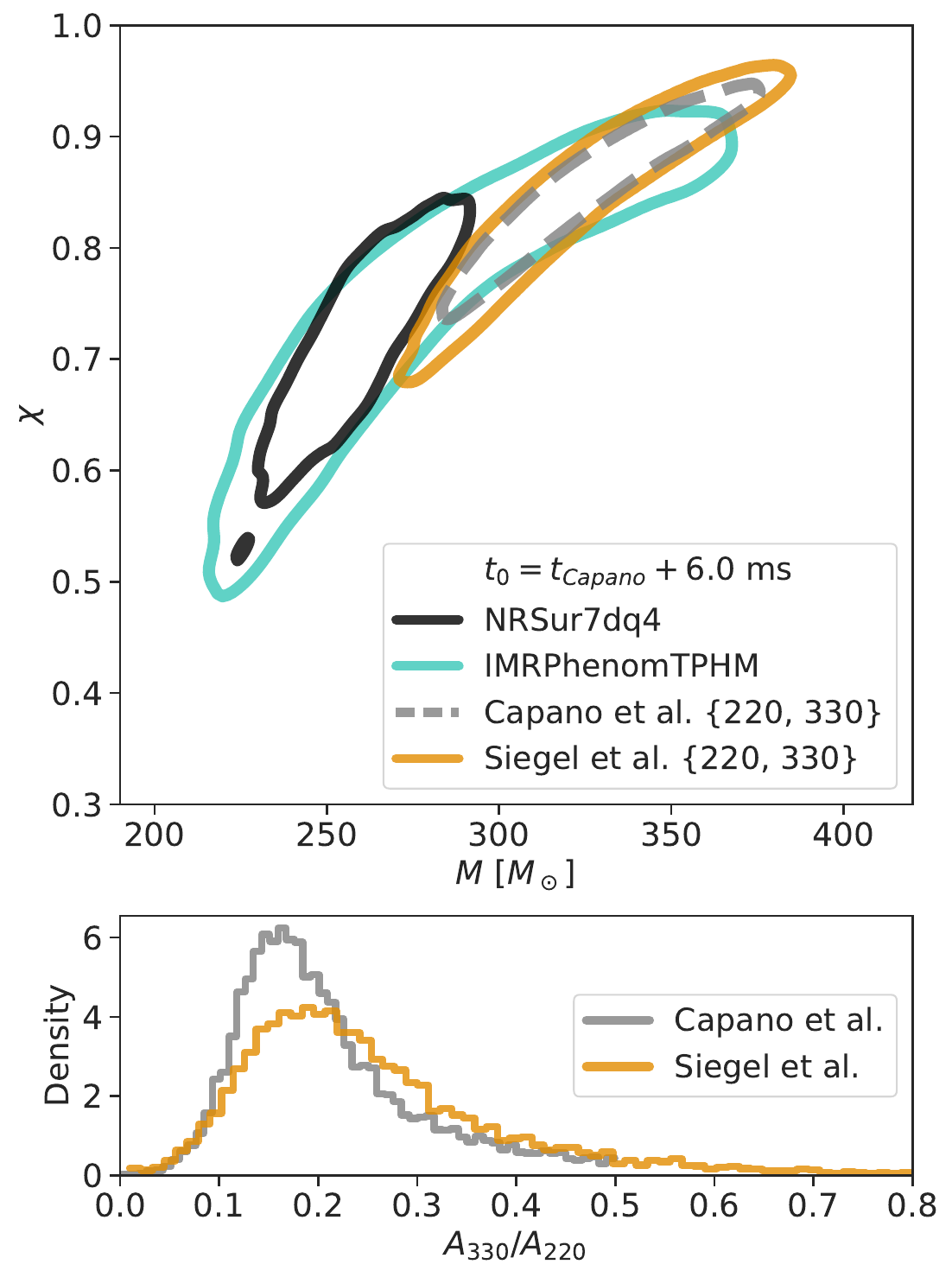}
    \caption{(Top) We attempt to reproduce the Capano \textit{et al.}~remnant mass and spin posterior shown in Fig.~3 of \cite{CapanoGW190521g_330} (dashed grey), using the same \{220,~330\} model (gold) starting 6.0 ms after their reported peak strain time. This start time is where Capano \textit{et al.} find that the Bayes factor for the \{220,~330\} model over the \{220\} or \{220,~221\} is maximized. We plot 90\% credible contours and find we can largely reproduce the posteriors of Capano \textit{et al.}, up to small differences owing to choices like their imposition of equatorial symmetry and QNM amplitude priors.
    We show IMR distributions from NRSur7dq4 (black) and IMRPhenomTPHM (teal, from data release of~\cite{Estelles_GW190521Waveforms}) for comparison.
    (Bottom) We also reproduce the amplitude ratio of the 330 and 220 modes.}
    \label{fig:FigCapanoCompare6ms}
\end{figure}

\begin{figure}
    {%
  \includegraphics[width=\columnwidth]{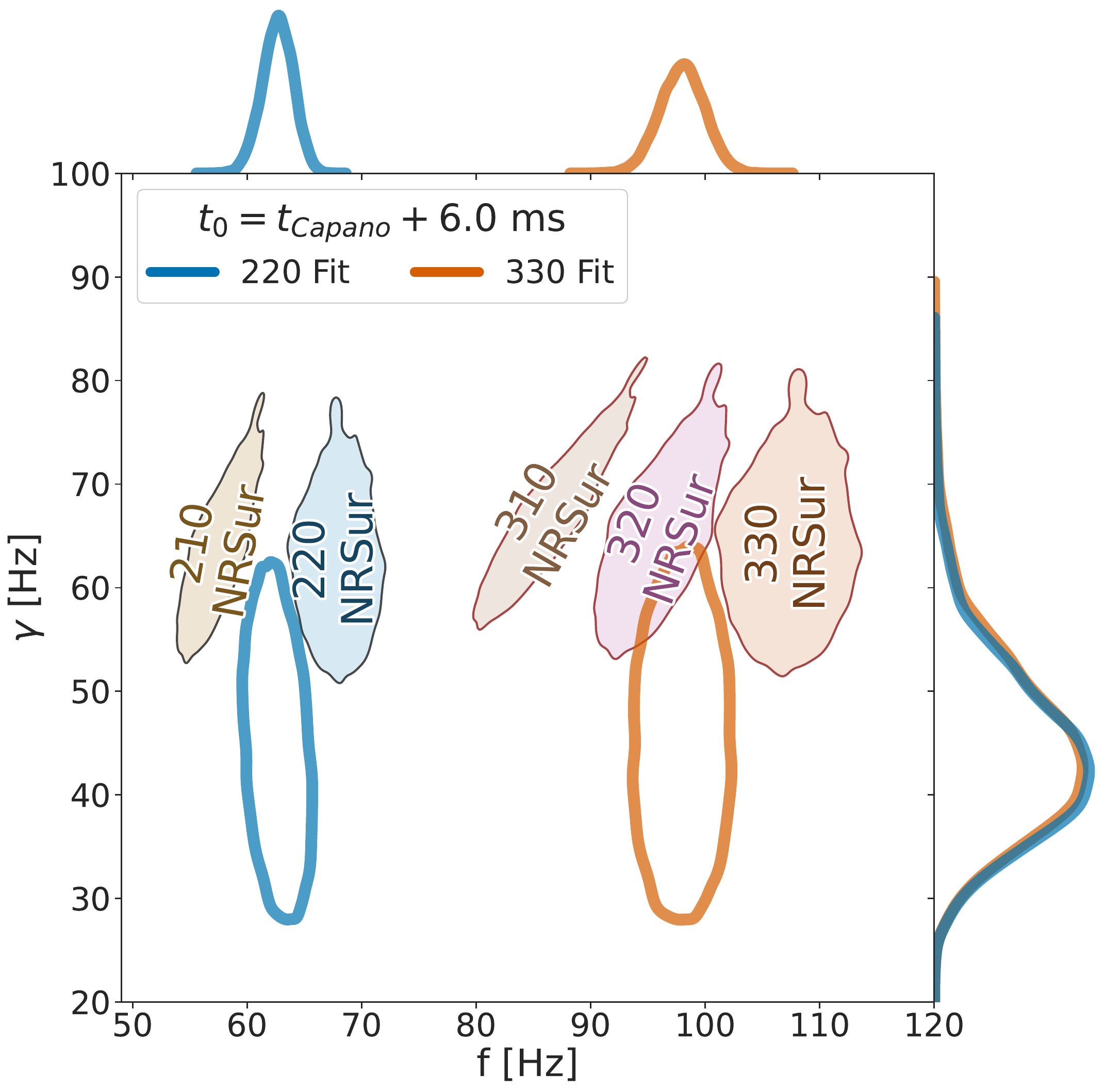}%
    }

    {%
  \includegraphics[width=\columnwidth]{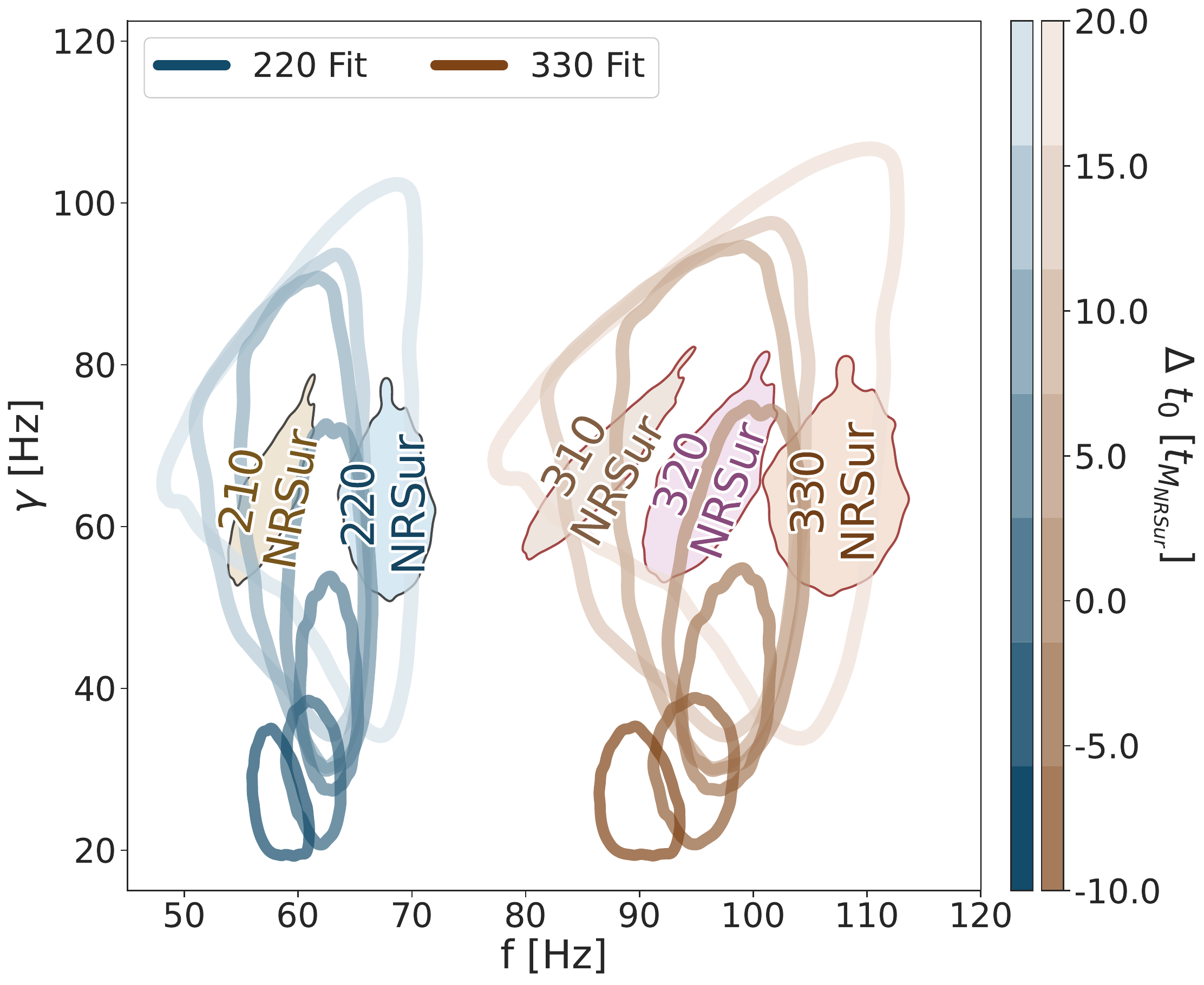}%
    }
    \caption{(Top) At the fit start time that Capano \textit{et al.}~claim produces the highest Bayes factor for the \{220,~330\} model over \{220\} or \{220,~221\} models, two features stand out when plotting QNM frequency and damping-rate: the frequency of our 220 fit lies between the 210 and 220 frequencies of NRSur7dq4, and the frequency of our 330 fit is more consistent with that of the NRSur7dq4 320. The same 220 behavior is present in Fig.~\ref{fig:220_fgamma}. (Bottom) The \{220,~330\} model and NRSur7dq4 are in tension until at least 20~$t_M$ after the peak, at which time the SNR has decreased substantially and uncertainties are large. All plots show  90\% credible contours.
    }
    \label{fig:220_330_fgamma}
\end{figure}

\subsection{Capano \textit{et al.}: \{220,~330\} Fit}

Capano \textit{et al.}~claim evidence of a higher-frequency subdominant QNM in GW190521. They report that this mode has a frequency of roughly 100 Hz and an amplitude around one order of magnitude smaller than that of the 220. Capano \textit{et al.}~propose that this subdominant QNM is the 330, and observe the largest Bayes factor in favor of a \{220,~330\} model compared to a \{220\} or \{220,~221\} model when fitting 6.0 ms after what they identify to be the time of peak strain (see Table \ref{tab:t0_compare}). Their analysis implements gating and in-painting \cite{Inpainting_Zackay} and performs fits in the frequency domain. Amplitude and phase tests of the Capano \textit{et al.}~posteriors \cite{FortezaRingdown}  show consistency with predictions from NR.

In Fig.~\ref{fig:FigCapanoCompare6ms} we reproduce the \{220,~330\} fit of Capano \textit{et al.}~6.0 ms after their reported peak time, and we find good agreement between our respective analyses when using this model. Notably, we confirm that the remnant mass and spin posterior of the \{220,~330\} model when fit at this time does not fully agree with NRSur7dq4.

The 330 mode could plausibly be excited to the extent reported by Capano \textit{et al.}~if the progenitors of GW190521 have an unequal mass ratio. The mass ratio posteriors given by most IMR waveforms tend to prefer an equal mass-ratio for GW190521, but there are also varying levels of support for unequal mass-ratio solutions depending on the waveform in question and the priors used \cite{Estelles_GW190521Waveforms,Olsen_GW190521_Likelihood, SimonaMiller_NRSurGW190521}.

Capano \textit{et al.} do find that the \{220,~330\} remnant mass and spin posterior overlaps with a region of parameter space where one IMR waveform, IMRPhenomTPHM \cite{IMRPhenomTPHM_Paper}, supports a $\sim$~4:1 mass ratio. However, the mass ratio and remnant mass posteriors obtained by IMRPhenomTPHM for GW190521 behave differently from the posteriors of most other IMR waveforms. NRSur7dq4 and IMRPhenomXPHM \cite{imrphenomxphm} both find the same detector-frame remnant mass of roughly $250~M_\odot$, largely independent of their mass ratio estimates. By contrast, IMRPhenomTPHM produces correlated estimates of the mass ratio and remnant mass such that at equal mass ratios the posterior prefers the remnant mass to be $\sim 250~M_\odot$, but at higher mass ratios the remnant mass increases to $300-350~M_\odot$, as shown in Fig.~2 of \cite{Capano_GW195021Validation}. These correlations may be evidence of waveform systematics.

In Fig.~\ref{fig:220_330_fgamma}, we plot the QNM frequency and damping rate posteriors of our \{220,~330\} fits. At the fit start time for which Capano \textit{et al.} find the highest Bayes factor in favor of the \{220,~330\} model, two aspects of the posteriors stand out: the frequency of the 220 fit lies between the 210 and 220 distributions inferred from NRSur7dq4, and the 330 fit has a frequency that is more consistent with that of the 320 mode of NRSur7dq4. Note that the 220 frequency and damping rate in this model behave in a similar way to the single-mode fits in Fig.~\ref{fig:220_fgamma}; the remnant mass and spin posteriors of the \{220,~330\} and \{220\} fits are nearly the same, which might be expected given the low amplitude of the 330.

\subsection{Summary of Previous Results}

We find that we can largely reproduce the results of both the LVC Kerr \{220\} and \{220,~221,~222\} models and the Capano \textit{et al.} \{220,~330\} model for GW190521. We also find that all of these analyses produce mass and spin posteriors which are not fully consistent with NRSur7dq4, over a large range of fit start times. In the case of \{220\} fits, we find that at later times the QNM frequency and damping rate become consistent with the 210 frequency and damping rate inferred from the NRSur7dq4 IMR fit. For \{220,~221,~222\} fits, the choice of QNM amplitude prior has a significant effect on the remnant mass and spin estimate; a fit that allows the overtone amplitudes to go to zero will produce posteriors in greater tension with NRSur7dq4. Finally, the frequency we measure for the 330 in our \{220,~330\} fits is more consistent with the 320 of NRSur7dq4. 

In Figs.~\ref{fig:Fig_LVC_10M_220_mchi},~\ref{fig:FigLVCCompare_0M},~and~\ref{fig:FigCapanoCompare6ms}, we have used similar data conditioning as was implemented by the LVC and Capano \textit{et al.} for the purposes of faithfully reproducing their results. Our own fits shown in Figs.~\ref{fig:220_fgamma}~and~\ref{fig:220_330_fgamma}, and in most of the other figures throughout this paper, use our own data conditioning choices unless otherwise indicated; see App.~\ref{sec:Appendix_AnalysisTechnicalChoices} for further details.

\begin{figure*}
  {%
    \includegraphics[width=\columnwidth]{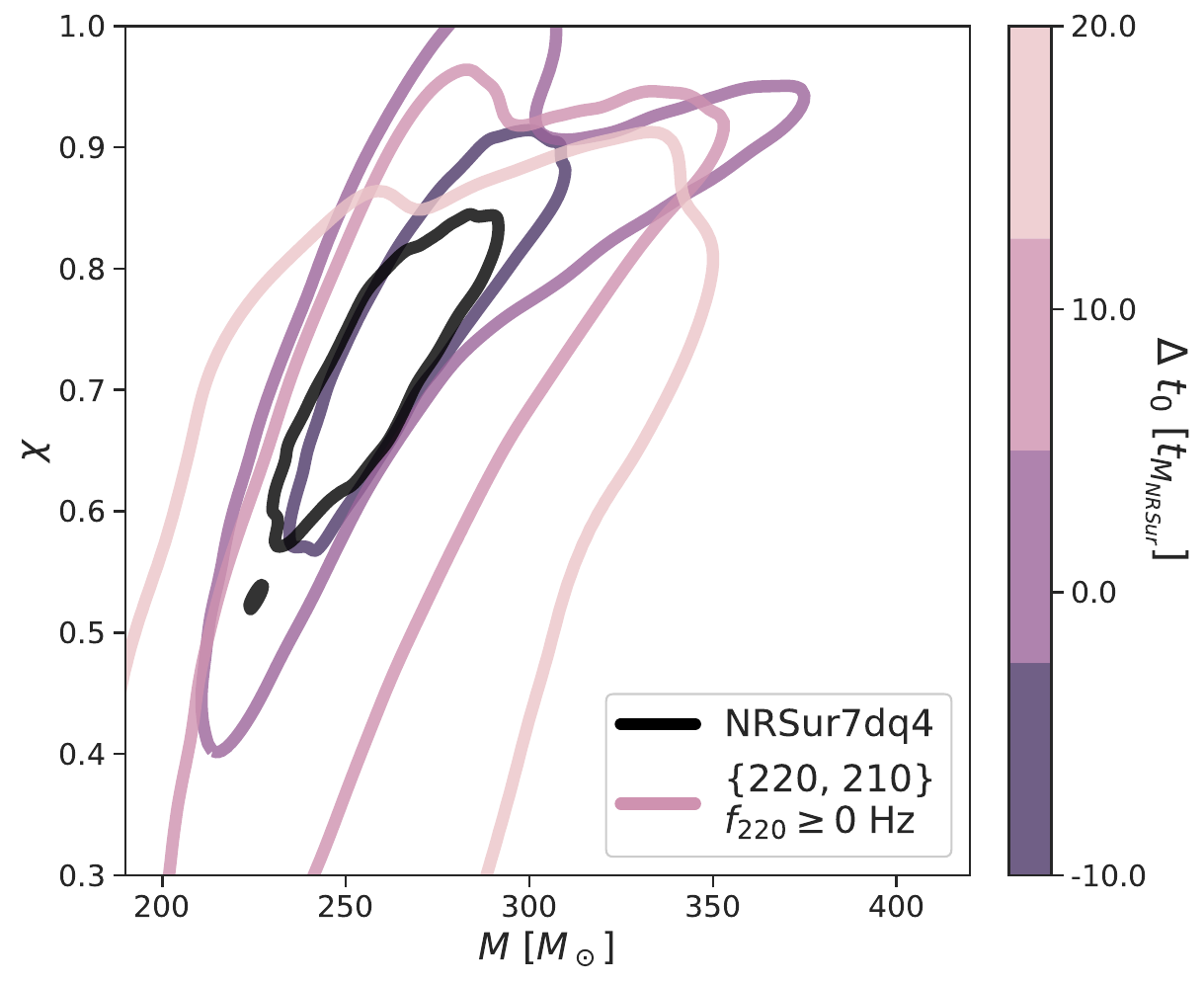}
  }\hfill
  {%
    \includegraphics[width=\columnwidth]{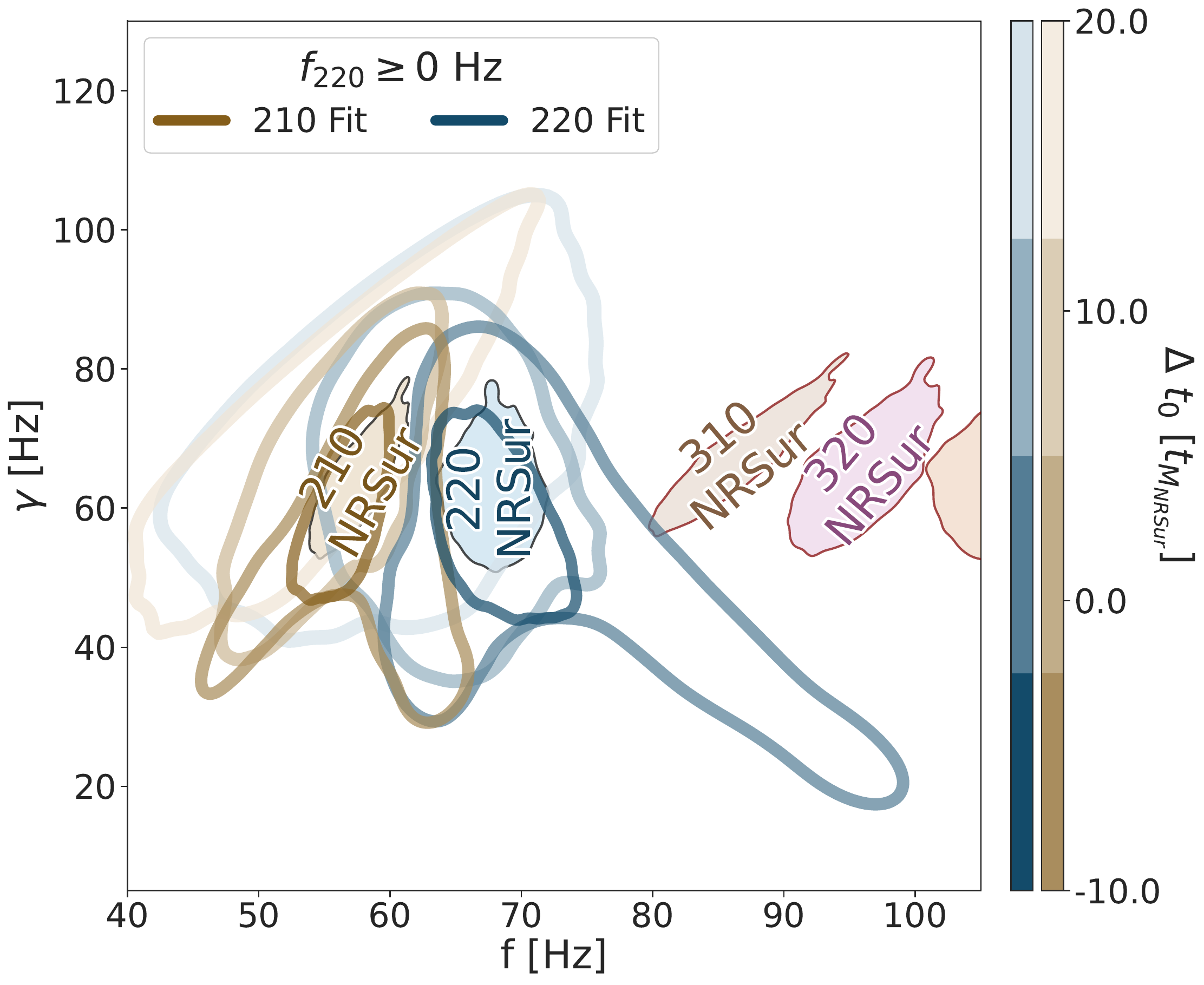}
}
    \caption{(Left) The \{220,~210\} model produces remnant mass and spin posteriors whose  90\% credible contours fully agree with those of NRSur7dq4 over a broad range of times. (Right) The same fits are shown in QNM frequency and damping-rate space. At early times, our fits of the 220 and 210 are in very close agreement with the distributions of the 220 and 210 given by NRSur7dq4. This early agreement may be related to the unusual lack of a prominent inspiral signal in GW190521, or the large peak strain timing uncertainty. At times near the peak strain, there are degenerate fit solutions: one solution agrees with NRSur7dq4, while the other solution places the 220 at higher frequencies and smaller damping rates. The higher frequency 220 fit corresponds to the location where the \{220,~330\} model places the 330 mode, once again consistent with the 320 frequency of NRSur7dq4. By 20 $t_M$, the posteriors begin reverting to the prior.}
    \label{fig:220_210_fgamma_and_mchi_timescans}
\end{figure*}

\section{Results: Siegel et al. }
\label{sec:Results}
In this section we explore ringdown models beyond those proposed by the LVC and Capano \textit{et al.}, in an effort to find a ringdown solution fully consistent with NRSur7dq4. Using the differences between the previous analyses' fits and the posteriors of NRSur7dq4 as motivation, we are first led to including the 210 mode in our fits. We find that \{220,~210\} ringdown fits for GW190521 give remnant mass and spin estimates in full agreement with those of most IMR waveforms, including NRSur7dq4, and these fits produce posteriors which are self-consistent over a range of fit start times. We then find that there is also some motivation for including a subdominant 320 mode in addition to the 210 mode.

We use a slightly different peak strain time estimate from those used by the LVC and Capano \textit{et al.}, see Table~\ref{tab:t0_compare}. We also implement flat amplitude priors on each QNM, a flat prior on the remnant mass from 0.5 to 2 times the median NRSur7dq4 estimate, and a flat prior from 0 to 1 on the remnant spin. For further technical information, see App.~\ref{sec:Appendix_AnalysisTechnicalChoices}.

\subsection{\label{sec:210_Fits}Siegel \textit{et al.}: \{220,~210\} Fit}
\begin{figure}
  {%
    \includegraphics[width=\columnwidth]{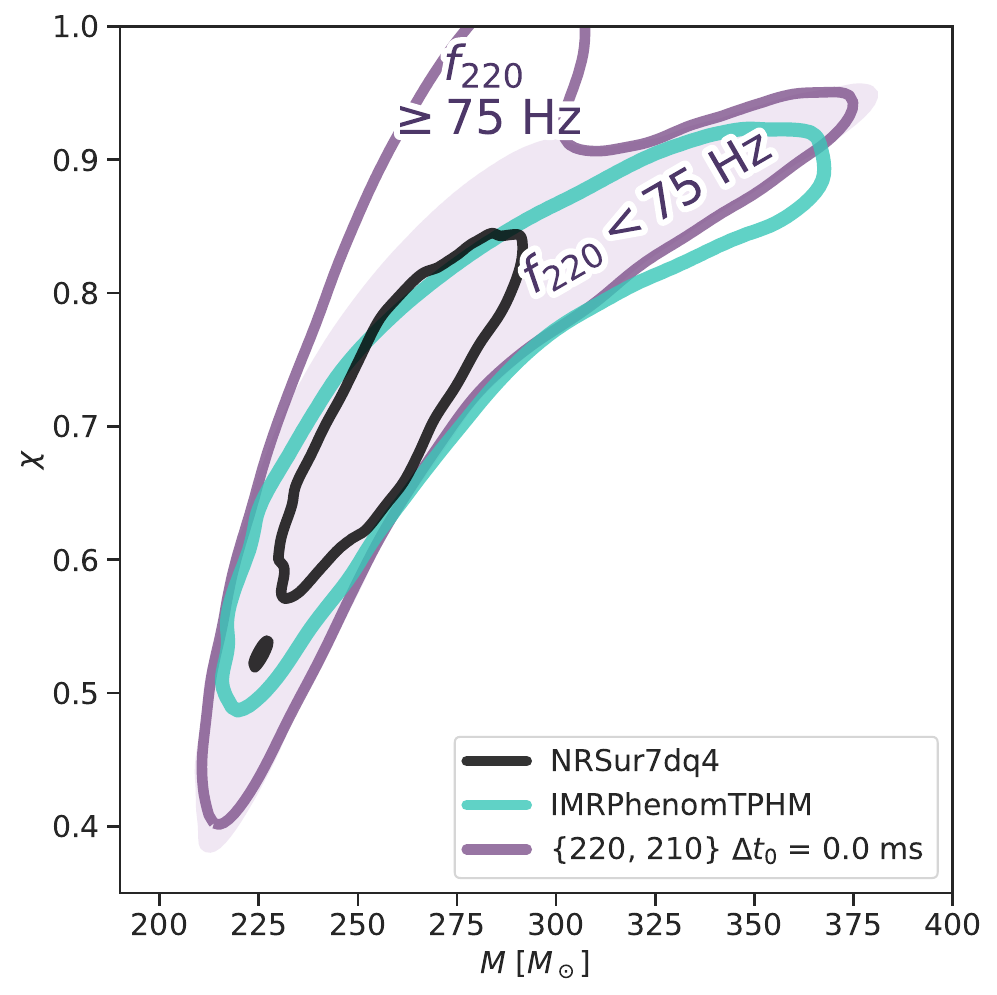}
  }
    \caption{Capano \textit{et al.}~claim that the IMRPhenomTPHM remnant mass and spin posterior is consistent with the \{220,~330\} model \cite{Capano_GW195021Validation}, although the \{220,~330\} posteriors are inconsistent with most other IMR waveforms. Here we show that the \{220,~210\} model when fit at the peak of the strain agrees with both NRSur7dq4 (black) and IMRPhenomTPHM (teal). We plot 90\% credible contours for two versions of the \{220,~210\} fit: the solid purple contour makes no QNM frequency cuts, whereas the shaded region contains only the degenerate low-frequency solution shown in Fig.~\ref{fig:220_210_fgamma_and_mchi_timescans}. The low-frequency solution simultaneously agrees with both IMRPhenomTPHM and NRSur7dq4 (see App.~\ref{sec:Appendix_PreviousAnalysesCont}). This level of agreement is not achieved by the \{220\} or \{220,~330\} models.} 
    \label{fig:220_210_mchi_0M_fmax75}
\end{figure}

\begin{figure}
  {%
    \includegraphics[width=\columnwidth]{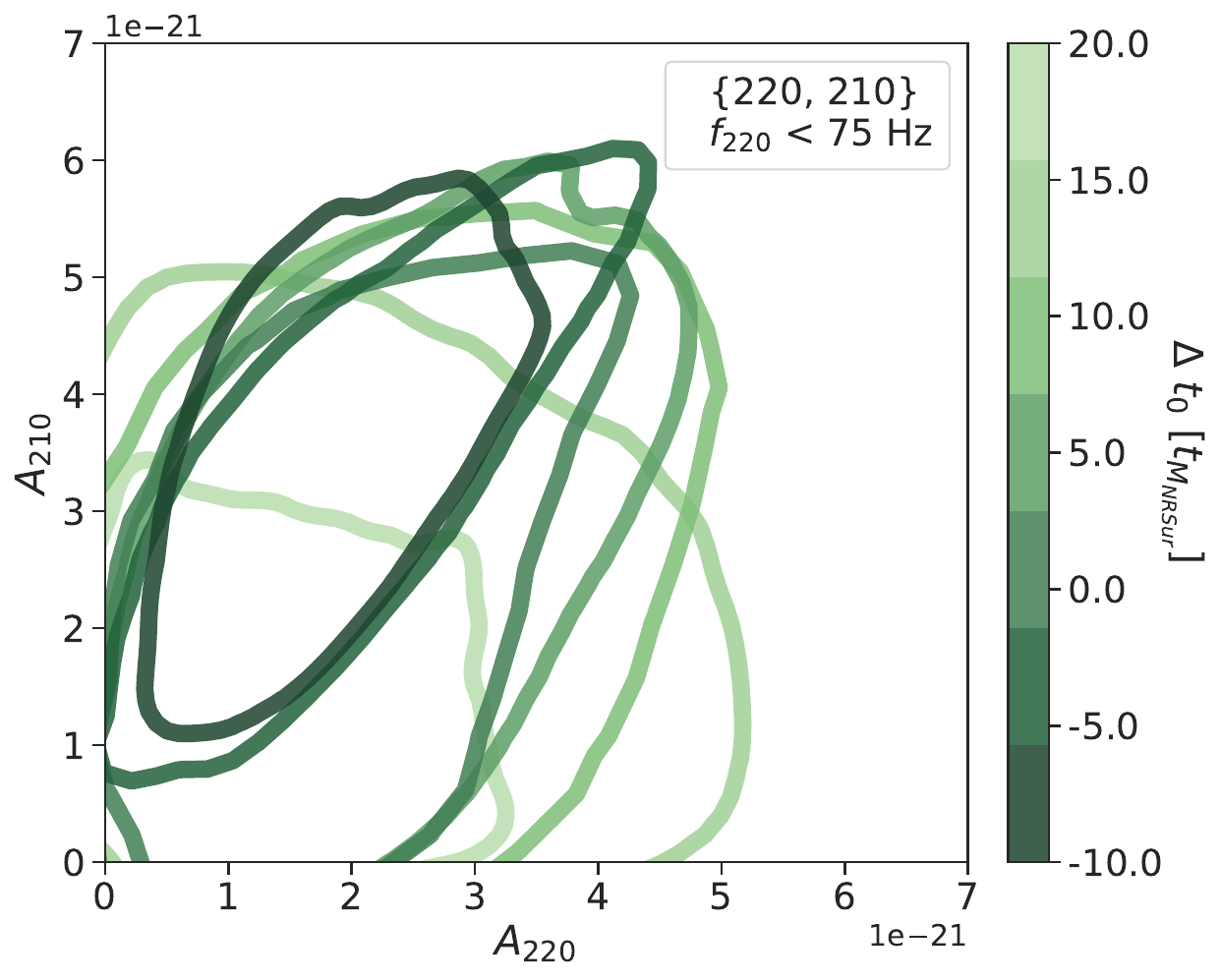}
  }
    \caption{Here we show 90\% credible contours of \{220,~210\} amplitudes over fit start time (green shades). Degenerate high-frequency solutions have been removed. The 210 amplitude (ordinate) is generally larger than the 220 amplitude (abscissa). This may explain the trends in Fig.~\ref{fig:220_fgamma}: we find that late-time single-mode fits tend towards the NRSur7dq4 210 parameters, which might be expected if the 210 has a larger amplitude than the 220 and is the last QNM to be obscured by noise.}
    \label{fig:220_210_A_timescan}
\end{figure}

As argued above and shown in Figs.~\ref{fig:220_fgamma}~and~\ref{fig:220_330_fgamma}, previous analyses' fits of the 220 QNM are not fully consistent with NRSur7dq4. We find that these inconsistencies are resolved by including the 210 QNM in our models.

In Fig.~\ref{fig:220_210_fgamma_and_mchi_timescans}, we show that remnant mass and spin posteriors from \{220,~210\} fits fully encompass those of NRSur7dq4 over a broad range of times. The posteriors of the \{220,~210\} fits at later times expand with increased uncertainty around the measurements from early-time fits, as opposed to shifting in parameter space over time, indicating that they are a robust description of the signal as it decays. 

However, our fits starting around the peak strain do contain degenerate solutions which disagree with the NRSurd7q4 parameters. These degeneracies can be seen more clearly in the QNM frequency and damping-rate plot of Fig.~\ref{fig:220_210_fgamma_and_mchi_timescans}, where the 220 posterior has a long excursion extending diagonally to the lower right, corresponding to higher frequencies and lower damping rates: this excursion places one of the fit modes at a frequency consistent with that of the 320 mode of NRSur7dq4 and places the other fit mode in between the 220 and 210 frequencies of NRSur7dq4.

In Fig.~\ref{fig:220_210_mchi_0M_fmax75} we focus on the remnant mass and spin distributions of the degenerate solutions around the peak time. Given that Capano \textit{et al.}~interpret the higher-frequency content in the signal as corresponding to the unequal mass-ratio part of the IMRPhenomTPHM posterior, we show that waveform's posterior for reference. By imposing a frequency cut at 75 Hz to select only the lower-frequency \{220,~210\} solution, we demonstrate that this solution has simultaneous agreement with both IMRPhenomTPHM and NRSur7dq4. This level of simultaneous agreement is not achieved by \{220\} or \{220,~330\} fits. We will implement this 75 Hz frequency cut for the remainder of this section. However, the higher frequency solution is still of interest as it indicates that there may be signal content at frequencies around 95 Hz, a prospect that we will explore further in the next section.

The \{220,~210\} remnant mass and spin posteriors closely overlap with those of NRSur7dq4 at early fit start times. The uncertainty of our peak time estimate is ${\sim\pm 2.5~ t_M}$. The remnant mass and spin posteriors almost fully agree even when fitting at ${\Delta t_0 = -5~t_M}$, meaning that the \{220,~210\} model gives consistent estimates of remnant mass and spin ${\sim 2 \sigma}$ before the median estimate of the peak strain time. Going further back than ${\Delta t_0 = -10~t_M}$, the ringdown posterior drifts away from the NRSur7dq4 distribution. The early-time agreement of our fits may be related to the lack of a prominent inspiral in GW190521, or the large peak time uncertainty for this signal.

Capano \textit{et al.}~find results similar to our early-time \{220,~210\} fits, although they attribute the observed behavior to noise. In Fig.~S.6 of~\cite{CapanoGW190521g_330}, for a non-Kerr \{220,~221\} fit that starts 7.0 ms before the reported peak time of this analysis, the overtone in the fit gets pulled by the free deviation parameters of the model towards the frequencies and damping rates of the 220 and 210 QNMs inferred by NRSur7dq4. Here the prior on the perturbed mode prevents its frequency from going below 55 Hz and thus cuts off a small portion of the 210 distribution.

In Fig.~\ref{fig:220_210_A_timescan} we show the evolution of the \{220,~210\} QNM amplitudes over time, restricting the posteriors to the low-frequency solution. The inferred amplitude of the 210 is comparable to, if not slightly larger than, that of the 220. This large 210 amplitude might explain the late-time trends shown in Fig.~\ref{fig:220_fgamma}, as it would imply that the 210 could conceivably be the only QNM detectable above the noise at late times, given that both the 220 and 210 modes have similar damping rates. 

Our flat amplitude prior allows each QNM's amplitude to go to zero; we find that imposing very strict frequency cuts such that the 220 and 210 can only have frequencies within the narrow ranges found by NRSur7dq4 results in amplitude distributions that are better constrained to be non-zero, for many fit start times. This suggests that the ringdown solution implied by NRSur7dq4 has significant contributions from both of these fundamental modes. See App.~\ref{sec:Appendix_RD_TheoreticalAnalysis_Details} for further technical discussion on the detectability of the 220 and 210.

Our findings regarding the 210 may be consistent with~\cite{Hoy_GW190521Precession}, which claims that GW190521 exhibits significant precessional power in the 21 spherical harmonic mode even more so than the 33 spherical harmonic mode.

\subsection{\label{sec:210_320_Fits}Siegel \textit{et al.}: \{220,~210,~320\} Fit}
\begin{figure}
    {%
  \includegraphics[width=0.99\columnwidth]{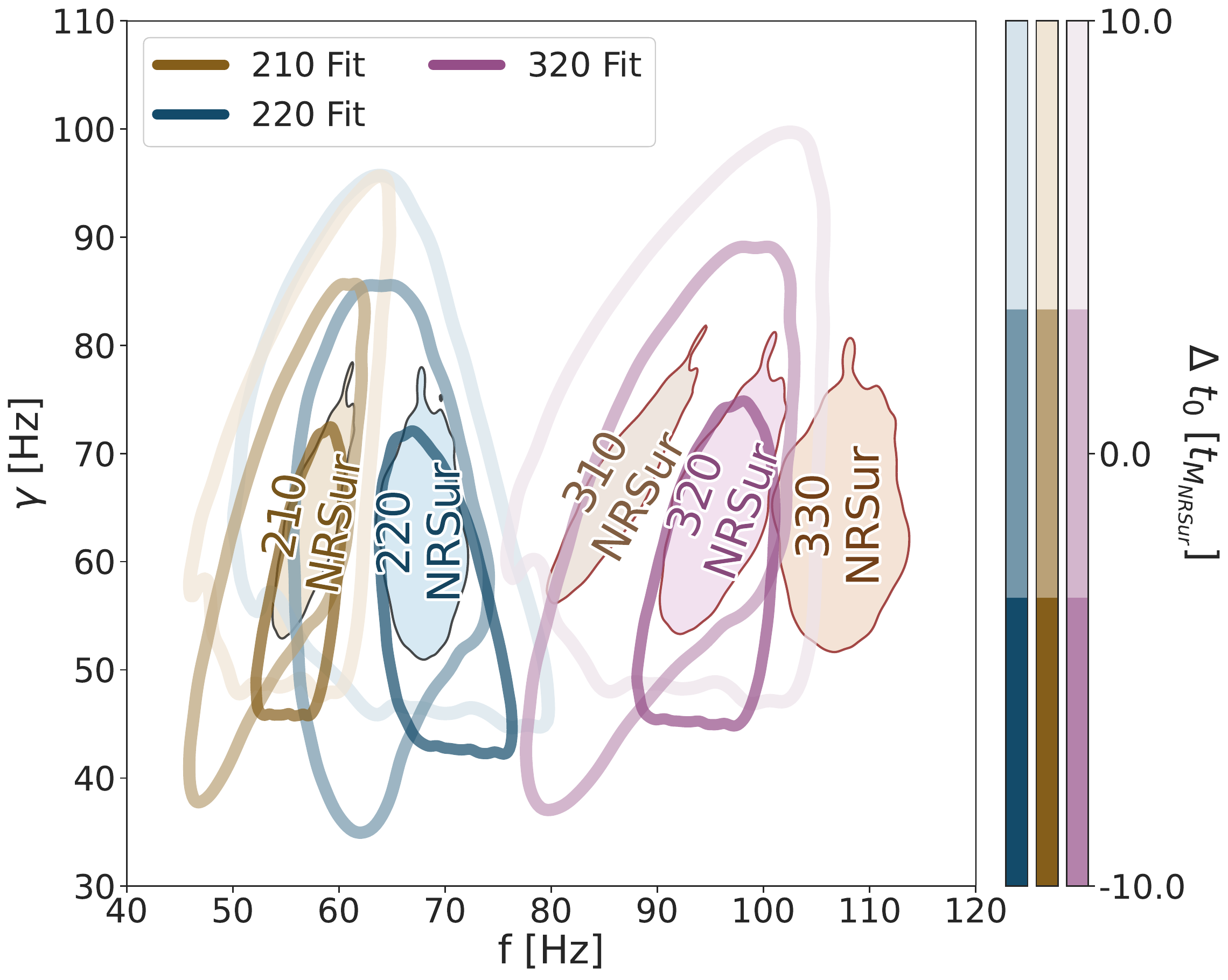}%
    }

    {%
  \includegraphics[width=\columnwidth]{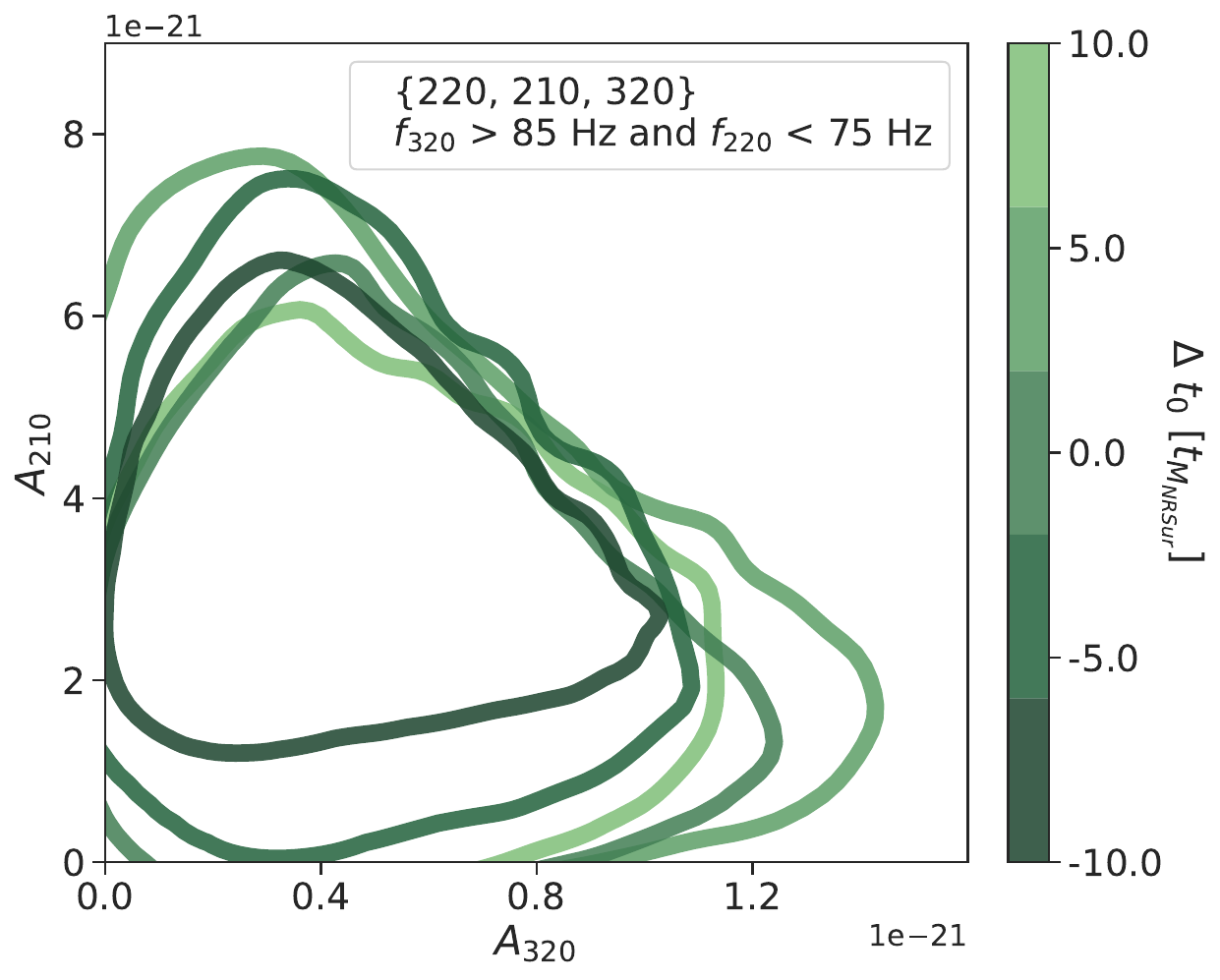}%
    }
    \cprotect\caption{(Top) QNM frequency and damping-rate of \{220,~210,~320\} fits over a range of fit start times. The posteriors of this model always contain a solution that fully agrees with NRSur7dq4. However, at later times two other degenerate solutions can be seen that are in disagreement with NRSur7dq4: one solution moves the 320 fit to the frequencies of the NRSur7dq4 ${l=2}$ QNMs, and another solution moves the 220 fit to the NRSur7dq4 ${l=3}$ frequencies as in Fig.~\ref{fig:220_210_fgamma_and_mchi_timescans}. (Bottom)~Amplitudes of the 210 (ordinate) and 320 (abscissa) QNMs from \{220,~210,~320\} fits. Frequency cuts are applied to eliminate label-switching from degenerate solutions, and to only consider solutions consistent with NRSur7dq4. The 320 amplitude is much smaller than those of the ${l=2}$ modes: note, this figure's axes are not scaled equally. The 210 and 220 amplitudes of this model are similar to what is shown in Fig.~\ref{fig:220_210_A_timescan}, and the correlations of the 320 and 220 are similar to what is shown here. 90\% credible contours shown.}
    \label{fig:320_Plots}
\end{figure}

Degeneracies in the fits of the last section suggest that there may be additional higher-frequency content in the data. Assuming that the high-frequency content is signal, we are guided by NRSur7dq4 to incorporate this content into our models as the 320 QNM.

In the top panel of Fig.~\ref{fig:320_Plots} we show \{220,~210,~320\} fits in QNM frequency and damping rate space over a range of times. At early times, the inferred QNM frequencies and damping rates agree almost completely with NRSur7dq4. Moving forward in time, the posteriors broaden while fully encompassing the NRSur7dq4 distributions. However, in addition to these IMR-consistent fits, two other degenerate fits also appear at later times. 

One of the degenerate solutions for the \{220,~210,~320\} model shifts all of the QNM frequencies downwards: the 320 is moved towards the 220 frequency of NRSur7dq4; the 220 is fit closer to the NRSur7qd4 210 frequency; and the amplitude of the 210 in our fits is zeroed out. This behavior is likely exacerbated by the fact that our uniform remnant mass and spin prior is so wide; in frequency and damping rate space, the prior that this imposes on the 320 QNM is peaked at frequencies below 75 Hz.

The second degenerate solution for the \{220,~210,~320\} model is similar to that seen in Fig.~\ref{fig:220_210_fgamma_and_mchi_timescans}, where the 220 is fit to higher frequency content around 95 Hz. If the signal of GW190521 had a much higher SNR, or our priors were more physically motivated in a way that enforced amplitude hierarchies for the modes in our fits, all of the degenerate solutions we have discussed might be avoided.

We impose frequency cuts to remove the degenerate solutions and select only those fits that are consistent with NRSur7dq4, and we then plot the inferred amplitudes of the 320 and 210 modes over time in the bottom panel of Fig.~\ref{fig:320_Plots}. We find that the ${l=3}$ mode is subdominant to the ${l=2}$ modes in our model, in agreement with Capano \textit{et al.} We also find that the amplitudes of both ${l=2}$ modes in this model are essentially the same as what is shown in Fig.~\ref{fig:220_210_A_timescan}. Since the \{220,~210,~320\} fits that agree with NRSur7dq4 do not have amplitudes consistent with zero for any of the three QNMs, we include all three modes in our non-Kerr analysis in Sec.~\ref{sec:TestsofGR}.

\subsection{\label{sec:Goodness_of_fit}{Quantifying Goodness of Fit}}
\begin{figure}
    {%
  \includegraphics[width=\columnwidth]{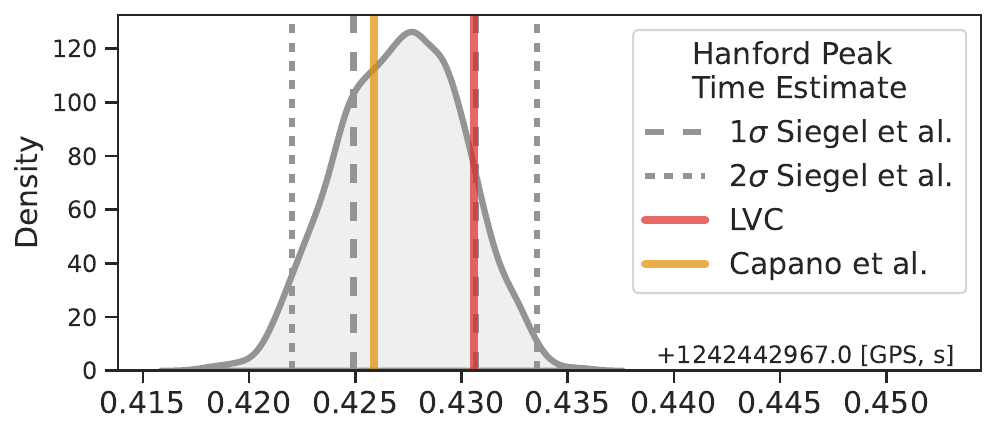}%
    }
    {%
  \includegraphics[width=\columnwidth]{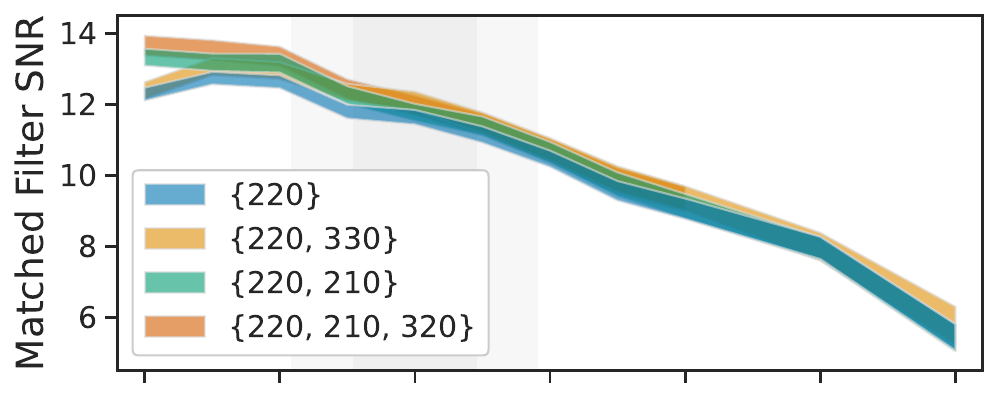}%
    }

    {%
  \includegraphics[width=\columnwidth]{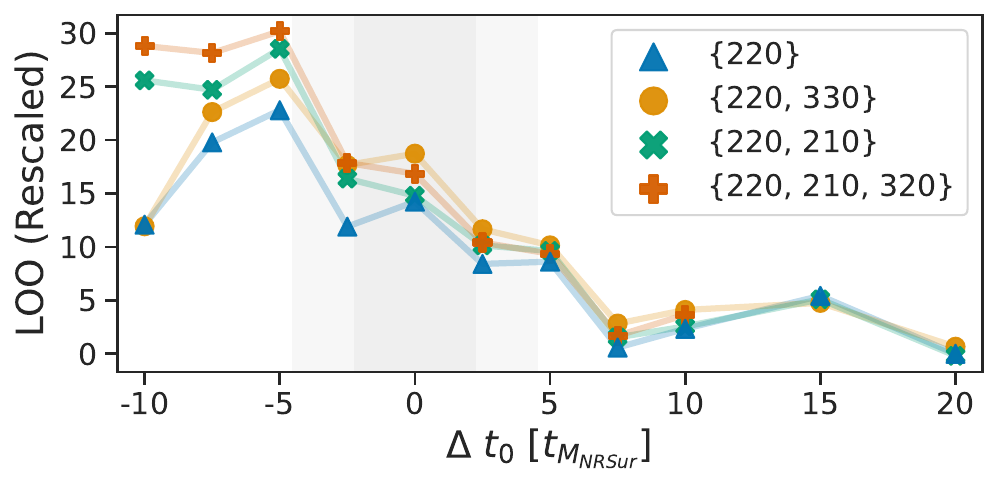}%
    }
    \cprotect\caption{(Top) Our peak strain time estimate (grey) in the LIGO Hanford detector, compared to reference times used by the LVC \cite{GW190521g_properties} (red) and Capano et al.~\cite{CapanoGW190521g_330} (gold). Also see Table~\ref{tab:t0_compare}. Timing uncertainties are shaded in the other two panels. (Middle) Network matched filter SNRs as a function of fit start time, 90\% credible intervals. We implement frequency cuts to avoid degenerate solutions (see Secs.~\ref{sec:210_Fits}~and~\ref{sec:210_320_Fits}). The SNR of all models starts to decay similarly after $-5\, t_M$. The SNR of the full IMR NRSur7dq4 fit is 14.6~$\pm$~0.4~\cite{GWOSC}. (Bottom) We use the leave-one-out cross-validation (LOO) to quantify data-driven goodness of fit for model comparison. Higher LOO values indicate better fits. We rescale the LOO such that its lowest absolute value over all times is zero. The trend of absolute LOO values over fit start time is only weakly informative, due to statistical uncertainty from the evolution of noise in each data segment. Still, the trends of LOO and SNR are strikingly similar. Using the \verb|compare| method in the \textsc{arviz} package \cite{arviz_2019} to estimate the significance of LOO differences, we find ${\sim 1 \sigma}$ preferences for \{220,~210,~320\}  at $-5\, t_M$ and for \{220,~330\} at 0 $t_M$, and ${\sim 1-2\sigma}$ preferences for all models over \{220\} at the same early times. After the peak, there are no strong preferences.}
    \label{fig:SNRs_and_LOOs}
\end{figure}

Our primary aim in this paper is to find Kerr ringdown solutions that agree with NRSur7dq4. However, there is no guarantee that the data itself is best-described by an IMR-consistent ringdown model, especially if the IMR waveforms are not accurate. Here we assess whether there are data-driven ringdown model preferences, independent of physics interpretations. Of the models discussed so far, we only consider those whose inferred QNM amplitudes are most consistent with being non-zero, and we thus exclude \{220,~221,~222\}.

To start, a simple way to compare models is by computing their recovered matched filter SNRs. We show in the middle plot of Fig.~\ref{fig:SNRs_and_LOOs} that the SNRs of the models we have considered are all very similar. The SNR of the ringdown signal in GW190521, while amongst the highest observed so far, is still relatively low. Thus it is not surprising that many QNM combinations can capture most of the signal's power, especially since our models of damped sinusoids are fairly unconstrained and flexible. Interestingly, the SNR appears to begin consistently decaying around $-5\, t_M$. At this time, we find that our models with the 210 QNM produce remnant mass and spin posteriors that overlap closely with those of NRSur7dq4.

For statistical model comparison, we implement the leave-one-out cross-validation (LOO) \cite{LOO_Paper, LOO_FAQ}. Although Bayes factors \cite{BayesFactors_WorkflowTechniques, CalderonBustillo_shoehorn} are more commonly used for model selection in gravitational-wave data analysis, we have chosen to instead implement the LOO because it is in principle less sensitive to prior regions without posterior support than the Bayes factor is. In brief, the LOO estimates the expected log pointwise predictive density from some observed data points $y_1,...,y_n$ modeled as independent given model parameters $\theta$:
\begin{equation}
    \text{elpd}_\text{LOO} = \sum_{i=1}^n \text{log}~p(y_i|y_{-i}),
    \label{elpd_loo_eqn}
\end{equation}
where
\begin{equation}
    p(y_i|y_{-i}) = \int p(y_i|\theta)~p(\theta|y_{-i})~\text{d}\theta
\end{equation}
and $y_{-i}$ denotes the dataset in question with the $i^{\text{th}}$ element removed. This quantity gives a measure of how well a fit would predict any single data point in a signal given the other data points. We compute the LOO from the whitened residuals of our QNM fits: higher LOO values correspond to whitened residuals which are more consistent with being white Gaussian-distributed noise. For further discussion and extensions of this method, see \cite{LOO_FAQ,LOO_limitations_vehtarigelman} and \cite{LOOCV_exogenousregressors} respectively.

In the bottom plot of Fig.~\ref{fig:SNRs_and_LOOs} we show LOO values for different ringdown models as a function of fit start time. It is reasonable to expect both from the lack of distinction in SNRs and from the degeneracies in our QNM posteriors that there will be no strong statistical preference for any one model, and this is essentially what we find when comparing LOO values. We use the \texttt{compare} method in the \textsc{arviz} package \cite{arviz_2019} to estimate the significance of LOO differences.\footnote{For technical reasons, unless imposed during Monte Carlo sampling (see App.~\ref{sec:Appendix_AnalysisTechnicalChoices}) we do not use frequency cuts to remove degenerate solutions when computing the statistical significance of LOO differences. The absolute values of the LOO are not significantly affected by any frequency cuts we make after sampling.} This method indicates ${\sim}1 \sigma$ preferences both for \{220,~210,~320\}  at $-5\, t_M$ and for \{220,~330\} at 0 $t_M$ respectively, and ${\sim} 1{-}2\sigma$ preferences for all models over \{220\} at the same early times. After the peak, we find no strong preferences for any model.

\subsection{\label{sec:TestsofGR}Non-Kerr Fits: Test of General Relativity}

\begin{figure*}\centering
\subfloat[]{\label{a}\includegraphics[width=.5\linewidth]{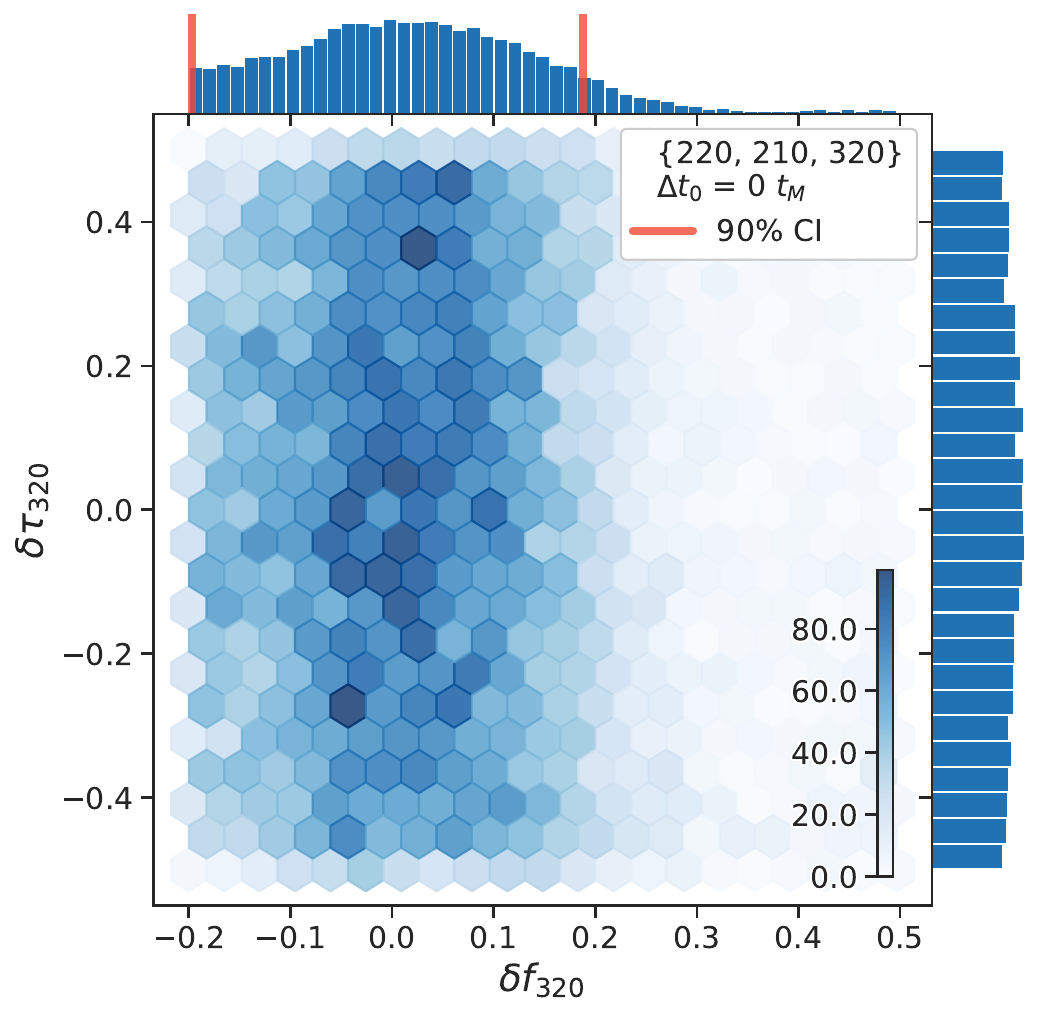}}\hfill
\subfloat[]{\label{b}\includegraphics[width=.5\linewidth]{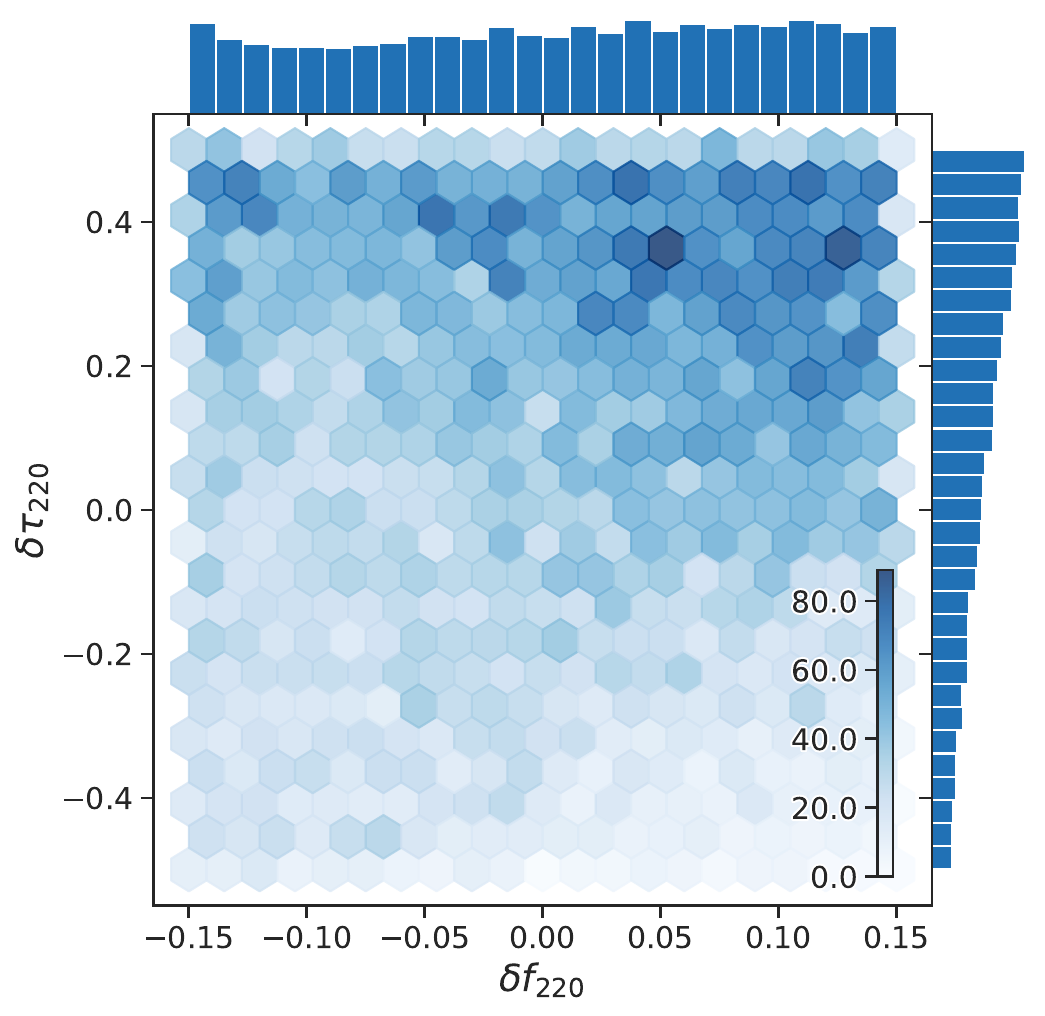}}\par 
\subfloat[]{\label{c}\includegraphics[width=.5\linewidth]{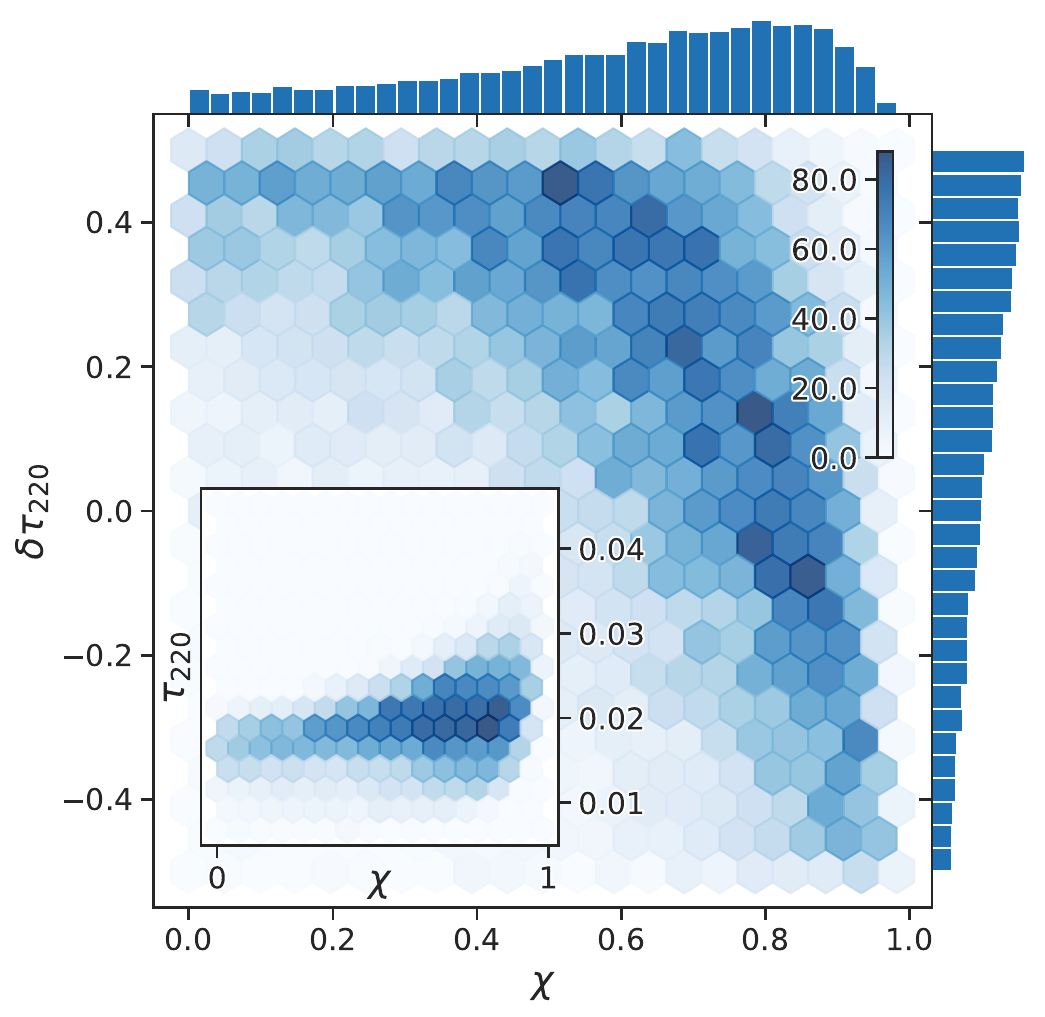}}
\caption{Non-Kerr fits of the \{220,~210,~320\} model at the peak strain time. Deviations are simultaneously allowed for the frequencies and damping times of the 220 and 320 QNMs. (A) We constrain $\delta f_{320}$ (abscissa) to within $\pm20\%$ at the 90\% highest-density credible level when fitting at 0~$t_M$. The lower end of this constraint is influenced by our prior. The constraint improves at $-5~t_M$, coming closer to $\pm10\%$ (not shown). (B) Our measurement of $\delta f_{220}$ (abscissa) is uninformative at 0~$t_M$. However, $\delta f_{220}$ is weakly peaked around 0 at $-5~t_M$ (not shown). As for $\delta \tau_{220}$ (ordinate), at first glance it seems to skew towards non-zero values, but this is the result of correlations with $\chi$. (C) Here we show that the skew of $\delta \tau_{220}$ (ordinate) is correlated with $\chi$ (abscissa). The correlations conspire to keep $\tau_{220}$ constant, as can be seen in the inset figure.}
\label{fig:TGR}
\end{figure*}

We have found the \{220,~210,~320\} Kerr model to give good agreement with NRSur7dq4. We can test the stability of fits with this model by introducing perturbations $\delta f$ and $\delta\tau$ to the frequencies and damping times of all but one QNM, such that ${f=f^{\text{GR}}(M,\chi)~\text{exp}(\delta f)}$ and ${\tau=\tau^{\text{GR}}(M,\chi)~\text{exp}(\delta\tau)}$. Non-zero values of $\delta f$ or $\delta\tau$ may signal a non-Kerr effect or model systematics. We choose to perturb the lowest-amplitude modes in our fits, the 220 and 320.

The prior placed on the perturbations needs to be chosen carefully to avoid mode switching and ensuing degeneracies \cite{Buscicchio:2019rir}. To this end, we implement uniform priors on the perturbations which are not necessarily symmetric about zero, in order to allow for the largest possible perturbations without label-switching.

The degeneracies that already exist within our Kerr fits further complicate the QNM label-switching issues, especially since some degenerate solutions zero out some QNM amplitudes. Any measured deviations from the Kerr QNM frequencies and damping times are only meaningful for modes with non-vanishing amplitudes. Thus, in addition to imposing a frequency cut at 75 Hz, we also put a Gaussian amplitude prior on the 320 QNM to enforce a hierarchy of amplitudes in our non-Kerr fits, with the 320 tending to be subdominant, in order to keep every QNM near the portion of parameter space where every amplitude is most consistent with being non-zero; see App.~\ref{sec:Appendix_AnalysisTechnicalChoices} and our data release \cite{DataRelease}.

In Fig.~\ref{fig:TGR} we show the posteriors of Kerr-deviation parameters for fits at the peak of the strain. We recover a constraint of  $\pm20\%$ on $\delta f_{320}$ around zero at the 90\% highest-density credible level, which is similar to the constraint that Capano \textit{et al.}~report in their own subdominant mode analysis. We find that this constraint is further tightened when fitting at $-5\, t_M$ (not shown). Our constraint on $\delta f_{220}$ is less informative, although once again it is more peaked around 0 when fitting at $-5\, t_M$ (not shown). At first glance it looks like $\delta \tau_{220}$ is skewed towards large values. However, this is the result of correlations between $\chi$ and $\delta \tau_{220}$ that conspire to keep $\tau_{220}$ constant. This seems to be a generic feature of this beyond-Kerr parameterization; the same effect appears in Fig.~S.5 of ~\cite{CapanoGW190521g_330} and  Fig.~4 of \cite{IsiNoHair_GW150914}, and also occurs when perturbing \{220,~210\} models for GW190521. The astrophysical population of parameters like $\chi$ may need to be jointly modeled alongside the non-Kerr deviations to avoid erroneously finding deviations from general relativity when analyzing catalogues of ringdown signals \cite{Payne:2023}.

\subsection{\label{sec:further_fits}Further GW190521 Ringdown Explorations}

We have fit several other Kerr ringdown models to GW190521 beyond those shown above, and we briefly discuss these additional models here. We do not find evidence that would lead us to prefer any of these fits.

For both \{220,~210,~200\} and \{220,~200\} fits, we find that the solutions that agree with NRSur7dq4 have 200 amplitudes consistent with zero. The noise level at the frequencies where NRSur7dq4 places the 200 is several times larger than it is at the 220 frequencies, and thus even a 200 excitation as large as those of the other ${l=2}$ QNMs might be undetectable: in going from 55 to 40 Hz, the PSD rises by a factor of $\sim$5 in all interferometers, and there are larger peaks at 50 Hz in Hanford and Virgo.

We also investigated overtone fits via \{220,~210,~221\} and \{220,~210,~211\} models, and did not find a preference for non-zero overtone amplitudes. This seems to be consistent with our findings regarding \{220,~221,~222\} fits in Sec.~\ref{sec:222Fits}.

Lastly, we considered several fits with ${l>2}$ modes. We tried \{220,~210,~310\} and \{220,~210,~330\} fits, and found that they either moved the 210 and 220 frequencies away from the NRSur7dq4 values or otherwise did not support non-zero amplitudes for the $l=3$ modes. We also tried \{220,~320\} fits and found that their frequencies could either be similar to our \{220,~330\} fits or our \{220,~210\} fits, but regardless were not fully consistent with NRSur7dq4. Finally, we performed \{220,~210,~320,~440\} fits and found no preference for non-zero 440 amplitudes.

Throughout this paper, we confined ourselves mostly to three-mode fits due to the computational expense of models with more modes. New features have since been added to the \textsc{ringdown} package to address this expense. We do not entirely rule out all possible fits with more than three modes, but given the SNR of the signal it seems unlikely that more modes could be fit convincingly.

\section{\label{sec:Discussion} Discussion} 

\subsection*{Imprints of Precession and Eccentricity on Quasinormal Mode Spectra}

Our results presented in Sec.~\ref{sec:Results} indicate that GW190521 ringdown fits which include the 210 mode are consistent with the remnant mass and spin distributions of NRSur7dq4; we also find some support for the inclusion of a subdominant 320 mode in our fits. The amplitudes of the 210 and 220 QNMs in our fits appear to be comparable in magnitude, as shown in Fig.~\ref{fig:220_210_A_timescan}. The 210 is not expected to be excited to this extent in the ringdowns of non-precessing quasi-circular BBHs. However, we propose that a large 210 amplitude could be excited in the ringdowns of precessing BBH coalescences.

In precessing systems, the spin orientations of the progenitors allow for misalignment of the binary's orbital angular momentum vector $\mathbf{L}$ and the remnant's spin vector $\boldsymbol{{\chi}}$. Since the angular content of the remnant perturbations is carried over from the inspiral \cite{Kamaretsos_Ringdown,BorhanianRingdown} and is oriented with respect to $\mathbf{L}$, the projection of these perturbations onto the frame aligned with $\boldsymbol{{\chi}}$ (where the QNMs are defined) will involve a rotation when there is remnant spin misalignment. Furthermore, for both non-precessing and precessing quasi-circular binaries, Post-Newtonian (PN) approximations \cite{Blacnchet_PN,Boyle_Precessing, Will_PNApprox} indicate that the dominant modes in the inspiral when observed in the co-orbital frame are generally the quadrupolar $(2, \pm2)$ modes \cite{Boyle_Precessing}. Thus, one might expect that the angular mode content of the ringdown strain from a precessing system will look to first order like a rotated version of the QNM spectrum from a non-precessing system, with the angle of rotation being the remnant spin misalignment angle. Rotations of spherical harmonic strain are performed using Wigner D-matrices \cite{WignerDQuaternions_Reference,WignerEuler_Reference,FinchMoore_PrecessingRingdown,wigner2012group}, which mix the amplitudes of harmonics that share the same $l$:

\begin{equation}
    h^\prime_{lm^\prime} = \sum_m\mathfrak{D}_{m^\prime m}^l (\mathbf{R})h_{lm},
\end{equation}
where $\mathfrak{D}_{m^\prime m}^l(\mathbf{R})$ is the D-matrix for a rotation $\mathbf{R}$, and $h_{lm}$ is the strain. This suggests that the precession-induced misalignment of $\mathbf{L}$ and $\boldsymbol{{\chi}}$ could be accompanied by the excitation of $l \neq m$ QNMs in the ringdown, with ${l=2}$ QNMs generally being dominant. In support of this hypothesis, preliminary studies of BBHs in the SXS waveform catalogue \cite{SXS_catalogue} indicate that there are large correlations between ${l\neq m}$ QNM amplitudes and the angle of spin misalignment of the remnant \cite{Zhu_SpinMisalignment}, with the 210 and 220 amplitudes even being comparable in the most highly misaligned systems.

Multiple analyses find support for the source of GW190521 being a precessing BBH \cite{Biscoveanu_MeasureSpinsHeavyBBH,Olsen_GW190521_Likelihood,GW190521g_properties,GW190521g_DiscoveryPaper}. In particular, NRSur7dq4 prefers large in-plane progenitor spins. By contrast, the ringdown models of the LVC and Capano \textit{et al.}~are more likely compatible with non-precessing quasi-circular BBHs, which could explain why their fits are not fully consistent with NRSur7dq4. As argued above, our IMR-consistent ringdown models have a reasonable interpretation in the context of precession; nevertheless, we do not have sufficient evidence to rule out other interpretations of GW190521 such as those invoking eccentricity~\cite{Eccentric_Gayathri, Eccentric_RomeroShaw}, because it is unclear if the QNMs in our models could be equally excited in alternative dynamical scenarios. Most theoretical studies of QNM amplitudes from the ringdowns of BBH coalescences have been limited to non-precessing binaries with equal or unequal mass ratios \cite{Xiang_Ringdown,BorhanianRingdown,FortezaRingdown,LionelLondon_RD_NR_Model,Zertuche_MultimodeRingdown,London_modelingringdownbeyondfundqnms, headoncollision_nonspinning, Mitman:nonlinearities_in_ringdown}. Preliminary progress has been made towards understanding the effects of spin misalignment on the ringdown \cite{Kamaretsos_Ringdown,FinchMoore_PrecessingRingdown,Eleanor_RingdownFreqPrecessing,Hughes_LargeqRingdown, Zhu_SpinMisalignment}, but general models have yet to be developed for astrophysical ringdowns.

As we broaden our theoretical understanding of astrophysical QNM spectra, it may become possible to make both eccentricity and precession measurements by analyzing the ringdown alone. For example, large 200 amplitudes should be tied to high eccentricity in non-spinning systems, since head-on collisions (the limit of maximal eccentricity) of non-spinning equal-mass black holes are dominated by the 200~\cite{headoncollision_nonspinning}. Even non-detection of any QNMs associated with eccentricity or precession could be used to place an upper bound on their amplitudes, which in turn could set constraints on different dynamical scenarios. This type of ringdown-only inference would improve upon traditional methods in the literature for measuring eccentricity and precession, as it is often assumed that such measurements are only possible if several cycles of the pre-merger strain are observed~\cite{Romero-Shaw:Eccentricity_or_Precession,SimonaMiller_NRSurGW190521}.

\section{\label{sec:Conclusion} Conclusion}

We find that when we include both the 210 and 220 quasinormal modes in our ringdown models of GW190521, we achieve agreement between the remnant mass and spin posteriors of our ringdown fits and the full inspiral-merger-ringdown fits of NRSur7dq4 over a broad range of ringdown fit start times. The inferred parameters of the 220 and 210 modes are stable over time; the posteriors at later times expand around those of earlier times. We find that such inspiral-merger-ringdown waveform agreement and self-consistency over time is not achieved by previous GW190521 ringdown analyses, which were performed by the LVC and Capano \textit{et al.}

When we perform analyses with flat priors on the amplitudes of each quasinormal mode, we find that the amplitude of the 210 mode is comparable to, if not slightly larger than, that of the 220. While this is not expected in non-precessing systems, we propose that such a large 210 excitation could be produced in the ringdown of a precessing system \cite{Zhu_SpinMisalignment}. This adds to a growing body of research which suggests that substantial information about binary progenitors may be encoded in quasinormal mode spectra \cite{Kamaretsos_Ringdown, Xiang_Ringdown,BorhanianRingdown, Hughes_LargeqRingdown}.

At fitting times near the peak strain, a degenerate solution can be found in our \{220,~210\} posteriors that seems to fit content in the data at frequencies of $\sim$95 Hz. This is the same frequency range where Capano \textit{et al.} claim to find a 330 quasinormal mode. Guided by agreement with NRSur7dq4, we interpret this mode to be the 320. We recover amplitude posteriors that are not consistent with zero for the 320 quasinormal mode when fitting it alongside the 220 and 210, and including this ${l=3}$ mode in our fits does not significantly change our remnant mass and spin estimates. We do not find support for the significant presence of any other quasinormal modes in this signal for fits that are consistent with NRSur7dq4.

For models that include the 210, our fits closely agree with NRSur7dq4 even when fitting at times before our median peak strain time estimate. There may be technical reasons for this, related to the unusual lack of a prominent inspiral signal in GW190521 or the large peak strain timing uncertainty. We plan to further explore this early fit behavior in future work.

When perturbing the Kerr quasinormal mode frequencies and damping times of our \{220,~210,~320\} fit at the peak strain time, we find a $\sim\pm$20\% constraint with 90\% credibility around zero deviation for the 320 frequency. We do not find as tight of a constraint on the 220 frequency deviation. There is a moderate skew towards positive damping-time deviations for the 220, but this is the result of correlations with $\chi$. These correlations seem to be a generic feature of this beyond-Kerr parameterization, and indicate that the astrophysical population of parameters like $\chi$ will need to be modeled jointly with non-Kerr deviations to avoid biases in tests of general relativity at the population level \cite{Payne:2023}.

In order for ringdown analyses to provide viable tests of general relativity in the strong field regime, it is important that we fully understand the astrophysical quasinormal mode spectra we should expect to see in the LIGO-Virgo-KAGRA catalogue. Mismodeling of these spectra may lead to erroneous measurements of deviations from general relativity. Even if the correct subset of Kerr quasinormal modes is chosen for a fit, without physically-motivated prior constraints there can still be many degeneracies in the posteriors of these models. It may be advantageous for future ringdown analyses to move away from current models made up of small subsets of unconstrained quasinormal modes, and towards models that include as many modes as possible and tune physics priors to determine which modes are likely to be dominant.

\begin{acknowledgments}
We thank Eliot Finch, Collin Capano, Yi-Fan Wang, Julian Westerweck,  Aaron Zimmerman, Asad Hussain, Hengrui Zhu, Andrew Gelman, Aki Vehtari, Eric Thrane, Paul Lasky, Katerina Chatziioanou, Simona Miller, Yuri Levin, and Vishal Baibhav for insightful discussion.

This material is based upon work supported by NSF's LIGO Laboratory which is a major facility fully funded by the National Science Foundation.
This research has made use of data or software obtained from the Gravitational Wave Open Science Center (gwosc.org), a service of the LIGO Scientific Collaboration, the Virgo Collaboration, and KAGRA. This material is based upon work supported by NSF's LIGO Laboratory which is a major facility fully funded by the National Science Foundation, as well as the Science and Technology Facilities Council (STFC) of the United Kingdom, the Max-Planck-Society (MPS), and the State of Niedersachsen/Germany for support of the construction of Advanced LIGO and construction and operation of the GEO600 detector. Additional support for Advanced LIGO was provided by the Australian Research Council. Virgo is funded, through the European Gravitational Observatory (EGO), by the French Centre National de Recherche Scientifique (CNRS), the Italian Istituto Nazionale di Fisica Nucleare (INFN) and the Dutch Nikhef, with contributions by institutions from Belgium, Germany, Greece, Hungary, Ireland, Japan, Monaco, Poland, Portugal, Spain. KAGRA is supported by Ministry of Education, Culture, Sports, Science and Technology (MEXT), Japan Society for the Promotion of Science (JSPS) in Japan; National Research Foundation (NRF) and Ministry of Science and ICT (MSIT) in Korea; Academia Sinica (AS) and National Science and Technology Council (NSTC) in Taiwan.
This paper carries LIGO document number LIGO-P2300214.

The Flatiron Institute is a division of the Simons Foundation. H.S. is supported by Yuri Levin's Simons Investigator Award 827103.

\textit{Software:} \textsc{ringdown}~\cite{ringdown_code}, \textsc{arviz}~\cite{arviz_2019}, \textsc{pymc}~\cite{Pymc}, \textsc{seaborn}~\cite{Seaborn}, \textsc{matplotlib}~\cite{matplotlib}, \textsc{jupyter}~\cite{jupyter}, \textsc{numpy}~\cite{numpy}, \textsc{scipy}~\cite{scipy}, \textsc{qnm}~\cite{qnmpackage_Stein},  \textsc{disbatch}~\cite{disbatch}, \textsc{pandas}~\cite{pandas}, \textsc{Python3}~\cite{Python3}.

\end{acknowledgments}

\appendix

\section{\label{sec:Appendix_AnalysisTechnicalChoices} Analysis Technical Choices}

We analyze 0.4 second segments of data containing the ringdown signal from version 2 of the released strain \cite{GWOSC}, at a sample rate of 4096 Hz. We use a median Welch estimate for our noise model, computed from 256 s of data around the event. We high-pass filter at 10 Hz, and employ a digital anti-aliasing filter to zero out frequencies above the Nyquist frequency while avoiding the extended high-frequency corruption imposed by the roll-on of commonly used filters (e.g., Butterworth). We also artificially inflate the value of the PSD below 10 Hz to censor those frequencies in our likelihood calculations. 

We estimate the peak time of the strain using an invariant sum of all $lm$ modes and samples from NRSur7dq4, as detailed in Table~\ref{tab:t0_compare}. We use a flat prior on remnant mass and spin, and generally use flat priors on QNM amplitudes unless otherwise specified. We do not assume equatorial symmetry of the QNM excitations. When performing Hamiltonian Markov-Chain Monte Carlo (MCMC) sampling of our likelihood for models that have an ${l=3}$ mode, we employ a frequency constraint such that the higher-order mode frequencies cannot go below 75 Hz. This is to prevent our sampler from getting stuck in regions where the ${l=3}$ mode is fit to signal that is labeled as ${l=2}$ by NRSur7dq4. We also find that the degenerate solutions that our sampler gets stuck on often have lower likelihoods, which is further motivation for the frequency constraints. See our data release for more details \cite{DataRelease}. Other frequency cuts made throughout this paper are implemented after MCMC sampling.

Fundamental aspects of our analysis are described in \cite{AnalyzingBHRingdowns}. A future companion paper \cite{TimeDomain_conditioningpaper} will discuss further theoretical aspects of time-domain ringdown analysis and motivate the data conditioning choices adopted here.

\section{\label{sec:Appendix_PreviousAnalysesCont} Further Discussion on Previous GW190521 Ringdown Analyses and IMR Consistency}

\begin{figure}
    \centering
    \includegraphics[width = \columnwidth]{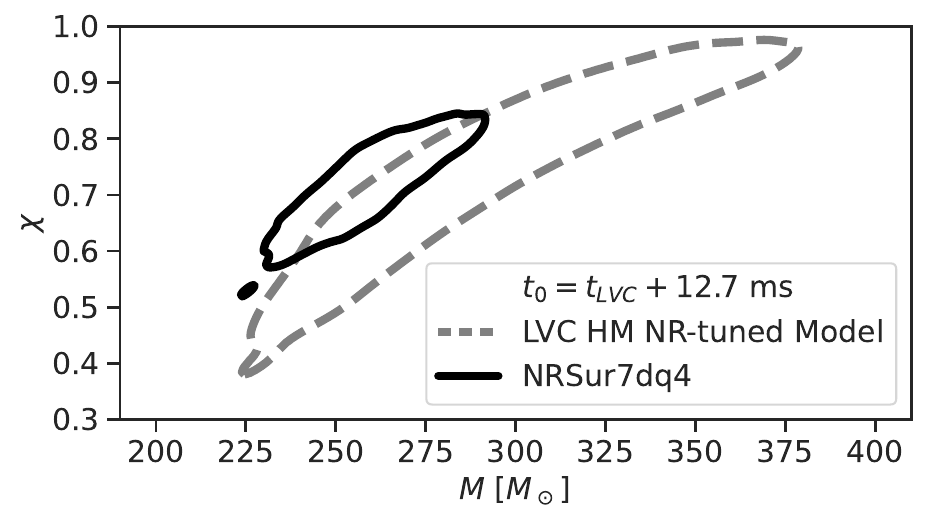}
    \caption{The LVC performed late-time fits using a higher-order angular mode model which is tuned to non-precessing quasi-circular NR systems. The 90$\%$ credible contour of this higher-order model's remnant mass and spin posterior (dashed) is inconsistent with that of NRSur7dq4  (solid), which may indicate that GW190521 is not well-described by non-precessing models.}
    \label{fig:LVC_Compare10M_HMNR}
\end{figure}

\begin{figure}
    \centering
    \includegraphics[width= \columnwidth]{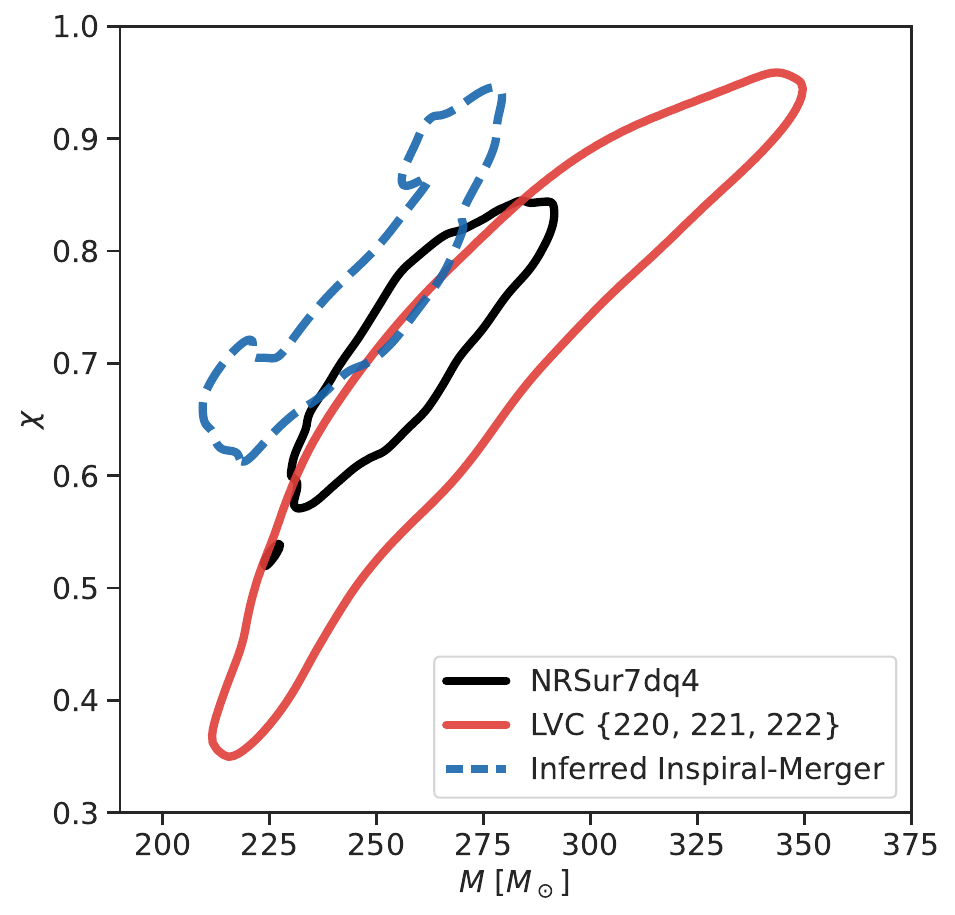}
    \caption{Using the assumption of Eq.~\eqref{Eq:Statistical_Independence_IM_R}, we demonstrate that seemingly small biases between ringdown (red) and IMR (black) posteriors may indicate mismodeling in at least one of the analyses. We plot 90$\%$ credible contours and show that if one were to assume that the two-overtone ringdown model shown here was consistent with the ringdown model preferred by NRSur7dq4, this would imply that remnant mass and spin estimates given separately by the inspiral-merger (blue dashed) of NRSur7dq4 and the two-overtone ringdown model are in disagreement.}
    \label{fig:IMR_consistency_demo}
\end{figure}

Here we discuss the level of agreement that should be expected between remnant parameter posteriors of IMR and ringdown analyses. It is reasonable to asssume that completely disjoint ringdown and IMR posteriors are in disagreement and that completely overlapping posteriors agree, but it is harder to define an acceptable amount of partial overlap of these posteriors. This problem is subtle: we contend that even partial overlaps such as those observed in Figs.~\ref{fig:FigLVCCompare_0M}~and~\ref{fig:LVC_Compare10M_HMNR} may be insufficient.

For an accurate IMR waveform, we would generally expect a ringdown model consistent with that IMR waveform to make similarly accurate but less precise parameter estimates. This means that the ringdown posteriors should fully encompass the IMR posteriors. Insufficient overlap of the ringdown and IMR posteriors can imply tension between the inspiral-merger portion of the waveform and the ringdown model in question. To demonstrate this point, assume that if one were to perform separate analyses of the inspiral-merger (IM) and ringdown (R) portions of a signal, they would produce two statistically independent measurements of the remnant mass and spin. That is, assume the likelihood functions for $M$ and $\chi$ can be multiplied to give a joint likelihood over the inspiral and ringdown phases of the signal:
\newcommand{\IM}{\mathrm{IM}}
\newcommand{\IMR}{\mathrm{IMR}}
\newcommand{\R}{\mathrm{R}}
\begin{multline}
  p\left( d_{\IM}, d_R \mid M, \chi, \IMR \right) = \\ p\left( d_{\IM} \mid M, \chi, \IM \right) p\left( d_{\R} \mid M, \chi, \R \right)
  \label{Eq:Statistical_Independence_IM_R}
\end{multline}
This assumption is, in practice, violated by the colored noise in current gravitational wave detectors: the likelihood for the inspiral is not independent of that for the ringdown. Nevertheless, the implications of this assumption are instructive, and we expect the correlations introduced by noise to be relatively small.

Given Eq.~\eqref{Eq:Statistical_Independence_IM_R}, partial overlap of remnant mass and spin posteriors obtained by separately analyzing the ringdown and IMR signals consequently necessitates partial overlap of posteriors obtained from the inspiral-merger and IMR signals. This can result in a significant lack of overlap between the posteriors from the inspiral-merger and ringdown, as shown in Fig.~\ref{fig:IMR_consistency_demo}. Inconsistencies like this are not expected if general relativity correctly describes the signals we observe, and thus lack of overlap of the IMR and ringdown analyses may indicate that at least one suffers from systematic inaccuracies.

Visually determining when sufficient overlap has been achieved is further complicated by nuances related to the spaces in which we plot our posteriors. The Jacobian used when transforming between remnant mass and spin space and QNM frequency and damping-rate space can alter the level of apparent disagreement between posteriors, making it harder in one of the spaces than the other to judge by eye the overlap of Gaussian kernel density estimates. For this paper, the QNM frequency and damping-rate emerged as more useful for qualitative assessments, both because they correspond more directly to observables in the data and also because trends like those in Fig.~\ref{fig:220_fgamma} may not be obvious without making comparisons to the individual modes given by IMR waveforms.

\section{\label{sec:Appendix_RD_TheoreticalAnalysis_Details} Detectability of Two Fundamental Modes with Nearly Equal Frequencies}

The 220 and 210 modes considered in this paper are quite close to each other in frequency and nearly identical in damping time.  The separation in angular frequency is $\delta \omega \simeq 10 \mathrm{Hz} \, \times 2 \pi$, which is comparable to their damping rate $\gamma \simeq 60 \mathrm{Hz}$.  It is reasonable to ask whether such modes are even detectable in principle, given the SNR of GW190521.  Consider a simplified model, where there are two (complex) modes with small angular frequency separation $\delta \omega$ and equal damping rates $\gamma$ in a signal,
\begin{equation}
  h(t) = \sqrt{2 \gamma} \left( A_1 e^{i \left( \omega + i\gamma \right) t} + A_2 e^{i \left( \omega + \delta \omega + i\gamma \right) t} \right),
\end{equation}
which is embedded in white noise with unit variance for $0 < t < \infty$.  The SNR of any signal $h$ in this situation is 
\begin{equation}
  \rho^2_h \equiv \int_{0}^\infty \mathrm{d} t \, \left| h(t) \right|^2;
\end{equation}
the normalization above is chosen so that $A_1$ is the SNR of mode 1 and $A_2$ the SNR of mode 2.  Assuming that the frequencies and damping rates of the modes are known perfectly, the Fisher matrix for the amplitudes is given by 
\begin{equation}
  F_{ij} \equiv \int_0^\infty \mathrm{d} t \, \frac{\partial h^*(t)}{\partial A_i} \frac{\partial h(t)}{\partial A_j}.
\end{equation}
In a measurement of mode amplitudes, the covariance matrix will be the inverse of this Fisher matrix:
\begin{equation}
  \Sigma \equiv F^{-1} = \begin{bmatrix}
    1 + 4 \xi^2 & -2 \xi \left( i + 2 \xi \right) \\
    2 \xi \left( i - 2 \xi \right) & 1 + 4 \xi^2 
  \end{bmatrix} ,
\end{equation}
where 
\begin{equation}
  \xi = \frac{\gamma}{\delta \omega}
\end{equation}
is the dimensionless ratio of the damping rate to the ``dephasing rate,'' or angular frequency separation.  

There are three interesting regimes.  When $\xi \ll 1$ then the modes are long-lived compared to their frequency separation, and the Fisher matrix approaches the identity, indicating that the amplitudes are measured with unit uncertainty (as expected for the measurement of an SNR) without correlation.  When $\xi \gg 1$ the modes damp rapidly compared to the rate at which they dephase, and the Fisher matrix becomes degenerate; the amplitude measurements are perfectly correlated, and this large degeneracy produces large uncertainty in the amplitude.  

Finally, when $\xi \simeq 1$ the modes are measurable, but with larger uncertainty, and the amplitude measurements are correlated with each other.  For GW190521, the 220 and 210 modes have $\xi \simeq 1.1$, which implies an uncertainty $\sigma_A \simeq \sqrt{1 + 4 \xi^2} \simeq 2.4$, or a significance of detection for a mode given by the mode's SNR divided by 2.4.  The correlation coefficient between the mode amplitudes is $\sim 0.9$.  Both of these features are observed in the posteriors we recover on the amplitudes of the 220 and 210 modes, shown in Fig.~\ref{fig:220_210_A_timescan}.



\bibliography{apssamp}

\end{document}